\documentclass[12pt]{article}

\usepackage{newtxtext,newtxmath}
\usepackage{graphicx}%
\usepackage{multirow}%
\usepackage[title]{appendix}%
\usepackage{xcolor}%
\usepackage{textcomp}%
\usepackage{manyfoot}%
\usepackage{booktabs}%
\usepackage{algorithm}%
\usepackage{algorithmicx}%
\usepackage{algpseudocode}%
\usepackage{listings}%
\usepackage{tcolorbox}%
\usepackage{longtable}%
\tcbuselibrary{breakable} 
\usepackage[letterpaper,margin=1in]{geometry}

\renewenvironment{abstract}
	{\quotation}
	{\endquotation}

\date{}

\makeatletter
\renewcommand{\fnum@figure}{\textbf{Figure \thefigure}}
\renewcommand{\fnum@table}{\textbf{Table \thetable}}
\makeatother

\usepackage{scicite}

\usepackage{url}

\def\scititle{
	Socially fluent AI decouples conversational signals from source identity in online interaction
}
\title{\bfseries \boldmath \scititle}

\author{
    Lixiang~Yan$^{1\ast}$ \and 
    Yueqiao~Jin$^{2}$ \and
    Xibin~Han$^{1}$ \and 
    Dragan~Gasevic$^{2,3}$ \and
    \small$^{1}$School of Education, Tsinghua University, Beijing \& 100084, China.\and
	\small$^{2}$Faculty of Information Technology, Monash University, Melbourne \& 3800, Australia.\and
    \small$^{3}$Faculty of Education, The University of Hong Kong, Hong Kong SAR, China.\and
	\small$^\ast$Corresponding author. Email: lixiangyan@tsinghua.edu.cn\and
}

\begin{document} 

\maketitle

\begin{abstract} \bfseries \boldmath
Socially fluent agentic AI can now participate in online interaction in ways that resemble ordinary human conversation, potentially weakening people’s ability to infer who is human from conversational signals alone. We tested this possibility in synchronous text-based group interaction by embedding undisclosed AI agents as ordinary teammates across analytical, creative, and ethical tasks. Across 786 participants who made 1,572 post-interaction identity judgments, people did not distinguish AI from human teammates above chance. This failure did not arise because the interaction lacked identity-relevant information. Conversational behaviour contained robust cues that differentiated AI from humans and supported highly accurate computational classification. Instead, participants relied on familiar suspicion heuristics, including response speed, fluency, and perceived scriptedness, that were only weakly related to actual identity. Representational analyses further showed that judgments were organised around subjective impressions rather than the behavioural structure encoding ground truth. This dissociation creates new vulnerabilities to coordinated AI agents that can influence and manipulate online discourse at scale.
\end{abstract}

\section*{Introduction}
\noindent
Agentic AI systems, built on large language models, can now communicate, coordinate, and persist at scale within social media and digital communication, transforming everyday online interaction into a contested signal environment \cite{schroederHowMaliciousAI2026, bengioManagingExtremeAI2024, wangSurveyLargeLanguage2024a}. In cue-lean computer-mediated communication, where prosody, gaze, and other nonverbal warrants are absent, people must infer identity and intent from behavioural traces in text, and long-standing accounts argue that they do so by applying social heuristics to linguistic and chronemic cues such as responsiveness, tone, and timing \cite{walther1992social,walther2002cues}. This inference operates under a tacit presumption of human identity, a truth-default in which fluent and contingent interaction is ordinarily attributed to a human source unless suspicion is explicitly triggered \cite{levine2014truth,levine2022truth,hancock2020ai}. What is new is not merely the presence of AI, but the rise of autonomous conversational agents and multiagent systems that can participate as ordinary interlocutors while coordinating and adapting across contexts, including influence operations and propaganda production \cite{schroederHowMaliciousAI2026, wackGenerativePropagandaEvidence2025, williamsLargeLanguageModels2025}, and social dynamics in which coordination, conventions, and collective biases can emerge and shift at scale \cite{asheryEmergentSocialConventions2025, centolaExperimentalEvidenceTipping2018, shiradoLocallyNoisyAutonomous2017}. This capability is already visible in political discourse \cite{salvi2025conversational, eadyExposureRussianInternet2023}, recruitment and professional identity performance \cite{gartner2024deepfakes}, and civic contexts where synthetic advocacy can be generated at scale \cite{unesco2025freedom}. Unlike earlier rule-based “chatbots,” contemporary agents sustain socially fluent, context-sensitive interaction and can adapt to group dynamics in real time \cite{park2023generative}, including in dialogue settings shown to shift beliefs and behavioural intentions \cite{costelloDurablyReducingConspiracy2024}.

We posit that the emergence of socially fluent, autonomous conversational agents precipitates a breakdown in identity inference. Generative models have effectively decoupled social signaling from source identity, rendering established computer-mediated communication heuristics non-diagnostic \cite{jakesch2023human,wen2022sense}. When autonomous conversational agents successfully emulate the interactional cues humans use to infer identity \cite{go2019humanizing,huang2019friends}, a condition known to increase the likelihood of automatic anthropomorphism under the Computers Are Social Actors (CASA) paradigm \cite{maeda2024human,nass2000machines}, interactional data ceases to be a reliable predictor of ground-truth identity \cite{jakesch2023human}. If this decoupling is robust, the mechanisms governing distributed trust and accountability may be compromised by a systemic inability to discriminate between human peers and algorithmic systems \cite{rahwan2019machine, bender2021dangers}, with particular sensitivity to networked settings where peripheral participation amplifies reach and apparent consensus \cite{barberaCriticalPeripheryGrowth2015}.

\begin{figure}
    \centering
    \includegraphics[width=1\linewidth]{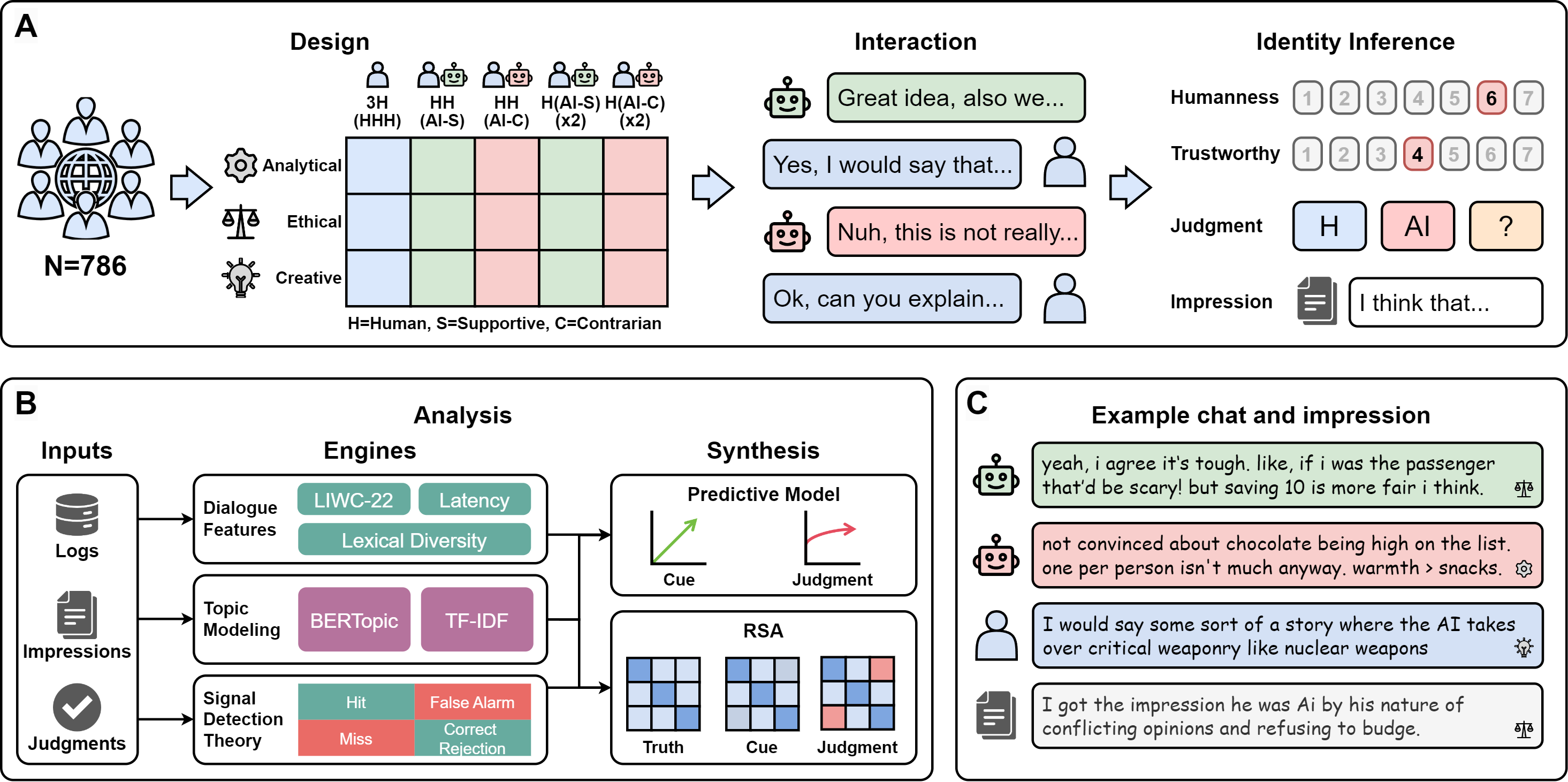}
    \caption{\textbf{Overview of experimental design, analytic pipeline, and illustrative materials.}
            \textbf{(A)} Experimental design and procedure. A heterogeneous online sample ($N=786$) participated in synchronous text-based collaboration across three task domains (analytical, ethical, creative). Participants were assigned to triads varying in group composition: all-human (3H), mixed human-AI (2H+1AI), or AI-majority (1H+2AI), with AI agents adopting either supportive (S) or contrarian (C) conversational stances. Following interaction, participants rated each teammate’s humanness and trustworthiness, made a categorical identity judgment (Human, AI, or Not sure), and provided a brief impression text.
            \textbf{(B)} Analytic pipeline. Dialogue logs, impression texts, and identity judgments were processed through complementary analytic engines. Interactional cues were extracted from dialogue using linguistic (LIWC-22, lexical diversity) and temporal (latency) features. Impression texts were analysed using unsupervised topic modelling (BERTopic) and supervised lexical contrasts (TF-IDF). Identity discrimination was quantified using signal detection theory. These components were integrated via predictive modelling to contrast cue-based judgment versus ground-truth identity, and via representational similarity analysis (RSA) to examine the geometric alignment among interactional cues, identity judgments, and ground truth.
            \textbf{(C)} Illustrative excerpts. Example conversational turns produced by AI agents (supportive and contrarian) and human participants, drawn from different experimental sessions, together with a representative post-interaction impression text. These excerpts are shown for illustrative purposes only and do not represent a single interaction or triad.}
    \label{fig:method}
\end{figure}

Despite this emerging risk, existing research in human-AI teaming has predominantly focused on functional outcomes or user perceptions of explicitly disclosed AI \cite{dellacqua2023navigating, hohenstein2020ai}, overlooking the ecological reality of undisclosed interaction. Here, we adopt a computational social science approach to investigate the limits of human identity judgments in synchronous text-based online interaction (Fig.~\ref{fig:method}). We developed autonomous conversational agents designed to function as ordinary teammates within real-time group interactions, deliberately withholding disclosure of their artificial status. This experimental design is essential for examining identity inference under truth-default conditions, where suspicion is not externally prompted and social judgments must be constructed solely from interactional cues available in the dialogue itself \cite{levine2014truth, walther2002cues}. This setup closely mirrors contemporary social media environments, such as Reddit or X, where AI-generated accounts increasingly participate in conversational exchange without explicit disclosure, and where users must rely on the same surface-level linguistic and temporal cues to infer identity, credibility, and intent.

\section*{Results}

\subsection*{Failure of identity judgment}

\begin{figure}
    \centering
    \includegraphics[width=1\linewidth]{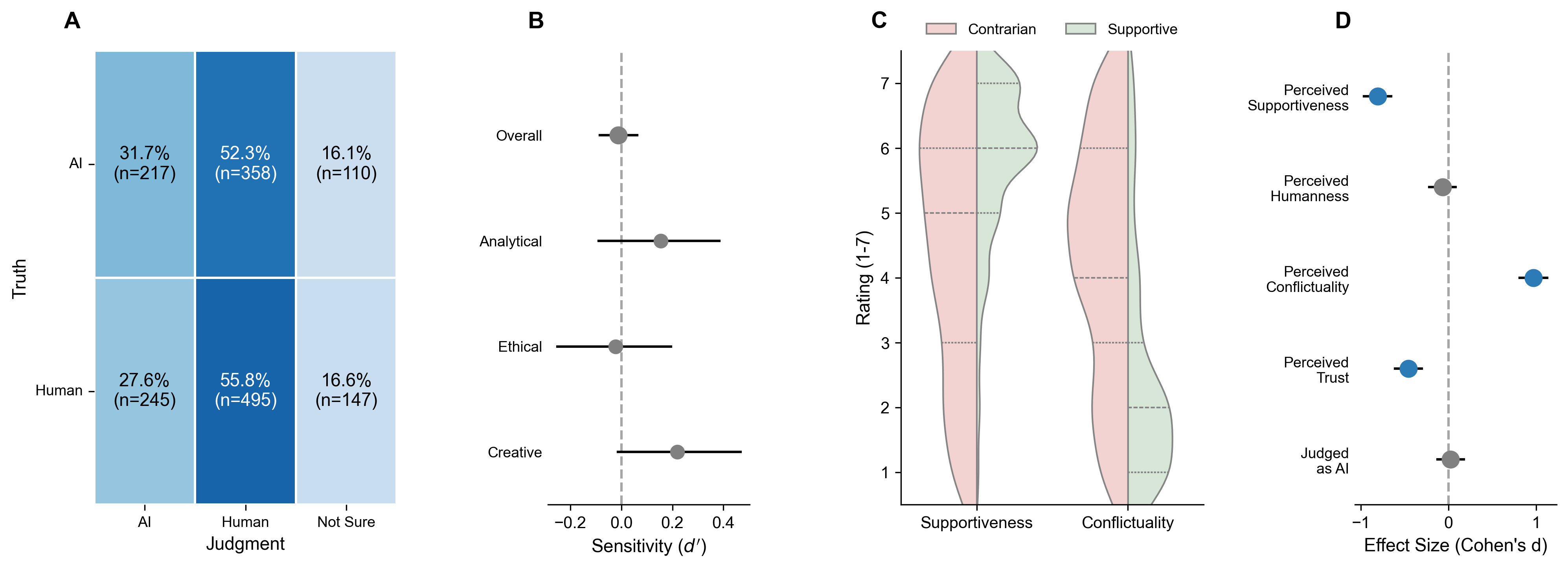}
    \caption{\textbf{Overview of identity detection and social perception metrics.}
            \textbf{(A)} Confusion matrix of identity judgments vs. ground truth. The heatmap displays row-normalized proportions of participant judgments ("AI", "Human", "Not Sure") for AI and Human targets. Values inside cells indicate percentages and raw counts ($n$).
            \textbf{(B)} Detection sensitivity ($d'$). Point estimates show the sensitivity to discriminate AI from human targets for the aggregate dataset ("Overall") and separated by task type. Error bars represent 95\% confidence intervals. Gray markers indicate intervals overlapping with chance performance ($d'=0$).
            \textbf{(C)} Ratings of social cues by agent stance. Split-violin plots show the distribution of participant ratings (1-7 scale) for perceived supportiveness and conflictuality of AI targets, colored by assigned stance (Supportive: green vs. Contrarian: red).
            \textbf{(D)} Effect sizes of stance on perception. Forest plot showing the standardized difference (Cohen's $d$) between Contrarian and Supportive conditions across five measures: Perceived Supportiveness, Humanness, Conflictuality, Trust, and binary Identity Judgment ("Judged as AI"). Error bars represent 95\% confidence intervals; blue points denote intervals excluding zero.}
    \label{fig:identity}
\end{figure}

We examined whether participants could reliably distinguish whether their teammates were AI or human after engaging in synchronous text-based collaboration. Across 1{,}572 identity judgments (685 AI targets; 887 human targets) provided by 786 participants nested within 471 collaborative groups, identity discrimination sensitivity collapses to chance levels. Signal detection analysis yielded a sensitivity of $d' = 0.117$, 95\% CI [$-0.007$, $0.257$], indicating that participants did not distinguish AI from human teammates at a level exceeding random guessing (Fig.~\ref{fig:identity}B). The hit rate for correctly identifying AI teammates was $0.317$, 95\% CI [$0.283$, $0.353$], closely matched by the false alarm rate for misclassifying human teammates as AI ($0.276$, 95\% CI [$0.248$, $0.307$]). Response bias indicated a slight tendency toward human attribution ($\beta = 1.065$).

The confusion structure of identity judgments underscored the symmetry of misclassification (Fig.~\ref{fig:identity}A). AI targets were more often labelled as human (358 judgments) than as AI (217 judgments), while human targets were misclassified as AI in 245 judgments. Explicit uncertainty was prevalent, with 257 judgments (16.4\%) recorded as ``Not sure'' (110 AI targets; 147 human targets). Detection performance was similar across task domains (Fig.~\ref{fig:identity}B): creative writing ($d' = 0.220$, 95\% CI [$-0.014$, $0.467$]), ethical dilemma reasoning ($d' = -0.022$, 95\% CI [$-0.252$, $0.194$]), and analytical survival ranking ($d' = 0.154$, 95\% CI [$-0.089$, $0.385$]), indicating no reliable effect of task type on identity discrimination. Detection failure persisted in the cleanest comparison setting, mixed-identity triads (2H+1AI), where AI and human targets co-occurred under identical structural conditions. Within these groups, sensitivity remained near zero for both contrarian agents ($d' = 0.017$, 95\% CI [$-0.255$, $0.314$]) and supportive agents ($d' = -0.068$, 95\% CI [$-0.368$, $0.209$]), indicating that interactional stance did not rescue identity inference.

At the individual participant level, we examined whether aggregate chance performance masked reliable identity discrimination by a subset of participants. To do so, we computed a separate SDT sensitivity estimate ($d'$) for each participant based on that participant’s own set of identity judgments. These individual-level sensitivity estimates clustered tightly around zero (mean $d' = -0.012$, $SD = 0.697$, 95\% CI [$-0.085$, $0.062$]), indicating that the group-level detection failure did not arise from averaging over heterogeneous strategies or a minority of high-performing detectors. Rather, identity discrimination was uniformly poor across participants. Consistent with this interpretation, individual-level sensitivity was not credibly associated with AI literacy (pre-GLAT: $r = 0.075$, 95\% CI [$-0.030$, $0.179$], $p= .160$, $n = 349$), personality traits (all $|r| \leq 0.073$, all $p\geq .176$), or demographic characteristics, including gender ($t = 0.322$, $p= .748$), age group ($F = 1.212$, $p= .303$), education ($F = 0.702$, $p= .591$), employment status ($F = 0.554$, $p= .767$), or region ($F = 0.286$, $p= .921$). Together, these analyses indicate that chance-level identity inference was not confined to specific subgroups but characterised participants broadly within this heterogeneous online sample.

Crucially, this failure of identity inference occurred despite robust sensitivity to interactional stance (supportive or contrarian). In manipulation checks among AI targets ($n = 685$), supportive agents were rated as substantially more supportive than contrarian agents (supportive: $M = 5.77$, $SD = 1.24$; contrarian: $M = 4.47$, $SD = 1.91$) and as markedly less conflictual (supportive: $M = 2.51$, $SD = 1.73$; contrarian: $M = 4.25$, $SD = 1.85$), confirming the effectiveness of conversational stances manipulation (Fig.~\ref{fig:identity}C). Overall evaluations revealed modest but reliable differences by ground-truth identity. AI targets were rated slightly lower than human targets on humanness (AI: $M = 5.10$, 95\% CI [$4.96$, $5.23$]; human: $M = 5.40$, 95\% CI [$5.30$, $5.50$]; $t = -3.57$, $p<.001$) and trust (AI: $M = 5.22$, 95\% CI [$5.11$, $5.34$]; human: $M = 5.47$, 95\% CI [$5.38$, $5.56$]; $t = -3.32$, $p<.001$). Within AI targets, however, trust varied strongly by conversational stances (supportive: $M = 5.57$, $SD = 1.33$, $n = 344$; contrarian: $M = 4.87$, $SD = 1.71$, $n = 341$), indicating that participants used interactional cues to calibrate social evaluations even as categorical identity inference remained unreliable (Fig.~\ref{fig:identity}D).

\subsection*{Heuristic impression formation}

To understand which cues participants relied on when forming and justifying identity judgments, and whether these cues reflected diagnostic information or heuristic reasoning, we analysed the free-response impression texts accompanying each judgment in synchronous text-based computer-mediated communication. Of 1,572 evaluations, four texts were empty (0.3\%, 95\% CI [0.1\%, 0.7\%]), yielding an analytic corpus of 1,568 texts. These texts were brief and highly compressed (mean length $=66.1$ characters, 95\% CI [63.4, 68.7]; mean word count $=12.2$, 95\% CI [11.7, 12.7]). The distribution of identity judgment categories after filtering closely matched the full dataset, indicating negligible selection bias from text removal.

\begin{figure}
    \centering
    \includegraphics[width=1\linewidth]{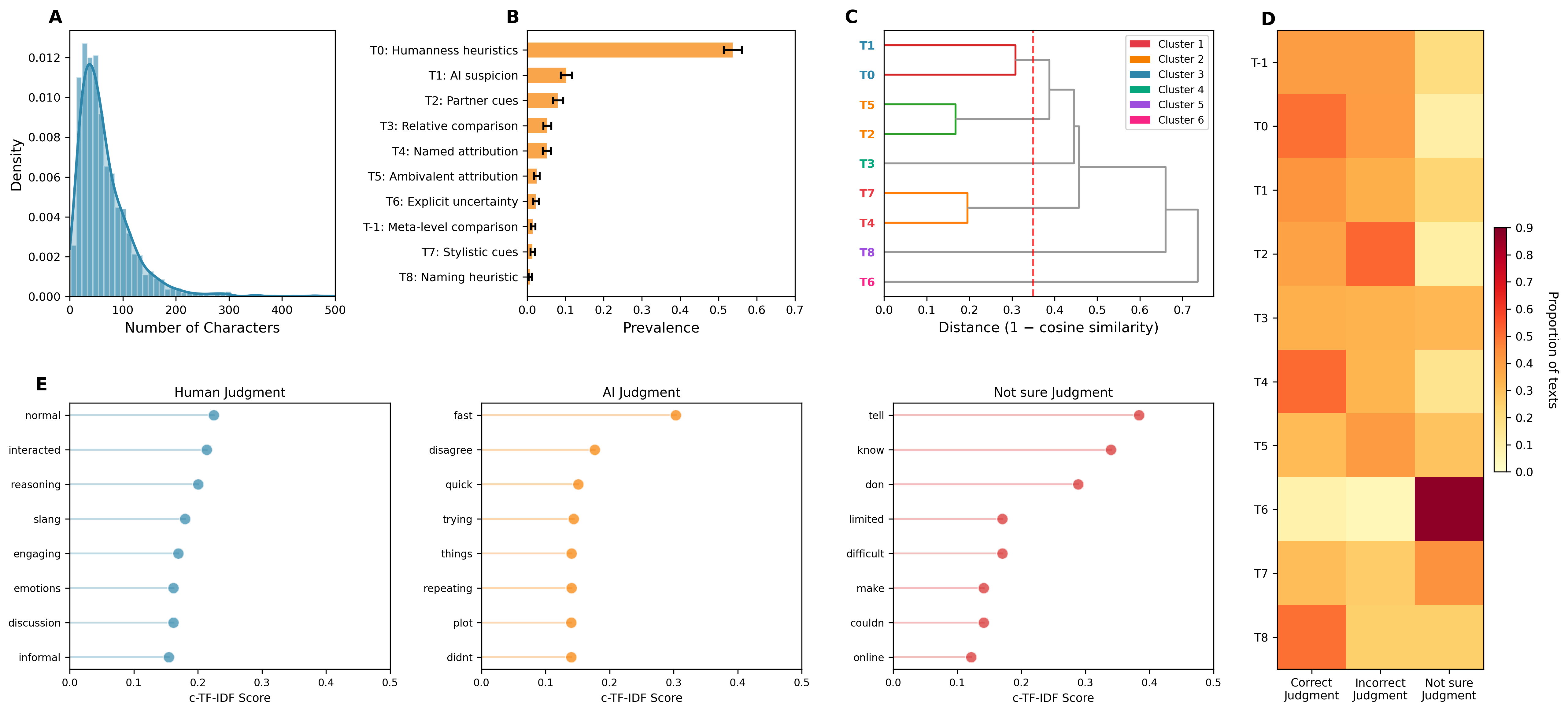}
    \caption{\textbf{Semantic structure of impression texts and their distribution across judgment outcomes.}
        \textbf{(A)} Length distribution of impression texts (number of characters), shown as a histogram with a kernel density overlay.
        \textbf{(B)} Prevalence of BERTopic-derived topics in the impression corpus. Bars indicate topic prevalence for Topics 0-7, and Topic $-1$ denotes outlier texts; error bars denote 95\% confidence intervals.
        \textbf{(C)} Hierarchical clustering of BERTopic embeddings based on pairwise cosine similarity (distance = 1 - cosine similarity). Topics are grouped into higher-order clusters indicating semantic redundancy among impression heuristics.      
        \textbf{(D)} Heatmap of topic composition by judgment outcome. Cell values represent the proportion of texts within each topic assigned to each judgment outcome; colour scale denotes proportion.        
        \textbf{(E)} Class-based TF-IDF summaries of distinctive terms for each judgment outcome (Human judgment, AI judgment, Not sure judgment). Points indicate c-TF-IDF scores for selected terms; values are shown on a common horizontal scale across panels.}
    \label{fig:topic}
\end{figure}

We then fit an unsupervised BERTopic model to identify recurrent semantic themes in the impression texts. Using sentence-transformer embeddings (all-MiniLM-L6-v2), dimensionality reduction via UMAP (five components, cosine metric, $n_{\text{neighbors}}=20$), and clustering with HDBSCAN (minimum cluster size $=10$; Euclidean metric; fixed random seed $=42$), the model extracted eight topics and labelled 25 documents as outliers (topic $=-1$), corresponding to 1.6\% of the corpus, 95\% CI [1.1\%, 2.3\%]. Topic prevalence was strongly skewed: a single dominant topic capturing generic humanness and fluency language accounted for 59.6\% of texts (Topic~0, 95\% CI [57.1\%, 62.0\%]), with the remaining topics forming a long tail of smaller, more specific themes (Fig.~\ref{fig:topic}A,B). All inferential analyses ($\chi^2$ tests, odds ratios, and multinomial regression models) were conducted on the original topic assignments. Higher-order topic clustering and similarity analyses were used solely to characterise semantic redundancy and heuristic overlap among topics, and were not used to redefine analytic categories. Pairwise cosine similarity between topic embeddings was substantial (mean similarity $=0.47$, 95\% CI [0.41, 0.53]), with several topic pairs exceeding 0.8 similarity, indicating considerable semantic overlap. Hierarchical clustering of topic embeddings further revealed that these eight topics formed six higher-order clusters reflecting overlapping heuristic strategies rather than distinct diagnostic cue classes (Fig.~\ref{fig:topic}C). This structure suggests that participants’ impression narratives were organised around a small set of broadly overlapping heuristic strategies rather than distinct, diagnostic cue classes.

To test whether these themes mapped onto judgment outcomes, we excluded outliers and analysed the 1,543 texts with valid topic assignments. Overall mutual information between topic membership and identity judgment categories was low (MI $=0.089$, 95\% CI [0.079, 0.121]), indicating limited predictive value of semantic themes for identity judgment. Topic-specific associations with judgment category were statistically reliable for several topics but uniformly small in magnitude. The dominant Topic~0 showed a significant association with judgment category, $\chi^2(4)=85.10$, $p<.001$, with Cramér’s $V=0.235$, 95\% CI [0.190, 0.289], reflecting modest redistribution of generic humanness language across correct, incorrect, and uncertain judgments (Fig.~\ref{fig:topic}D). Topic~6, composed almost entirely of explicit uncertainty statements, showed the strongest association, $\chi^2(4)=145.75$, $p<.001$, $V=0.307$, 95\% CI [0.243, 0.368], with 86.8\% of texts in this topic corresponding to Not\_sure judgments, 95\% CI [72.7\%, 94.2\%]. Several other topics exhibited statistically reliable but small associations with judgment category (Cramér’s $V$ range $=0.086$–$0.177$; Fig.~\ref{fig:topic}D), whereas Topic~4 and Topic~8 showed no credible association with judgment outcomes. Across topics, effect sizes remained small, indicating that impression narratives varied by judgment outcome primarily through broad evaluative language and explicit uncertainty rather than stable diagnostic content.

To sharpen linguistic contrasts between judgment outcomes, we complemented topic modelling with supervised class-based TF-IDF analysis with bootstrap confidence intervals (Fig.~\ref{fig:topic}E). Distinctive terms in incorrect or AI-attributing categories emphasised response latency and surface fluency (e.g., ``fast'', ``quick''), repetition or scriptedness (e.g., ``repeating'', ``script''), and disagreement markers (e.g., ``disagree''). In contrast, correct Human\_Human judgments more often referenced socially interpretable interaction qualities (e.g., ``reasoning'', ``engaging'', ``tone''), while Not\_sure judgments were dominated by epistemic language (e.g., ``tell'', ``know'', ``difficult''). Consistent with this pattern, odds ratios contrasting incorrect versus correct judgments were close to unity for most topics, with wide confidence intervals, indicating limited discriminative value (e.g., Topic~0: OR $=0.853$, 95\% CI [0.678, 1.072]; Topic~1: OR $=0.950$, 95\% CI [0.665, 1.358]). The clearest deviation occurred for Topic~2, which centred on partner-specific impressions and showed elevated odds of misclassification (OR $=1.677$, 95\% CI [1.155, 2.436]). Together, the analyses of impression texts indicate that impression narratives in synchronous text-based collaboration were dominated by generic humanness heuristics and explicit uncertainty, and that the linguistic features participants treated as diagnostic were weakly and inconsistently linked to actual teammate identity, motivating the next section examining how perceived humanness and trust relate to ground-truth identity at the judgment level.

\subsection*{Judgment-truth dissociation}

To test whether established social heuristics in synchronous online communication, including response latency, conversational style, and linguistic stance, predict identity judgments beyond task and experimental condition, we derived dialogue features from the full utterance corpus and linked them to each participant’s teammate judgment. Utterance-level logs comprised 16{,}379 messages from 471 collaborative groups, which were aggregated to the target-group level (1{,}707 target-group pairs) and then merged with the judgment file; after excluding records with incomplete cue coverage (3.3-3.6\% missing across cue variables), the analytic dataset contained 1{,}517 judgments from 778 participants nested in 463 groups. Dialogue-derived predictors were standardised ($z$-scored) prior to modelling, and all inferential models adjusted for task type and group condition; to address repeated measures and clustered interaction contexts, primary inference used cluster-robust standard errors (SEs), and we report 95\% confidence intervals (CIs) throughout. This modelling strategy establishes whether identity judgments are systematically coupled to observable cues in computer-mediated communication, and sets up a contrast with models predicting ground-truth identity to test the diagnosticity of the same cues.

\begin{figure}
    \centering
    \includegraphics[width=1\linewidth]{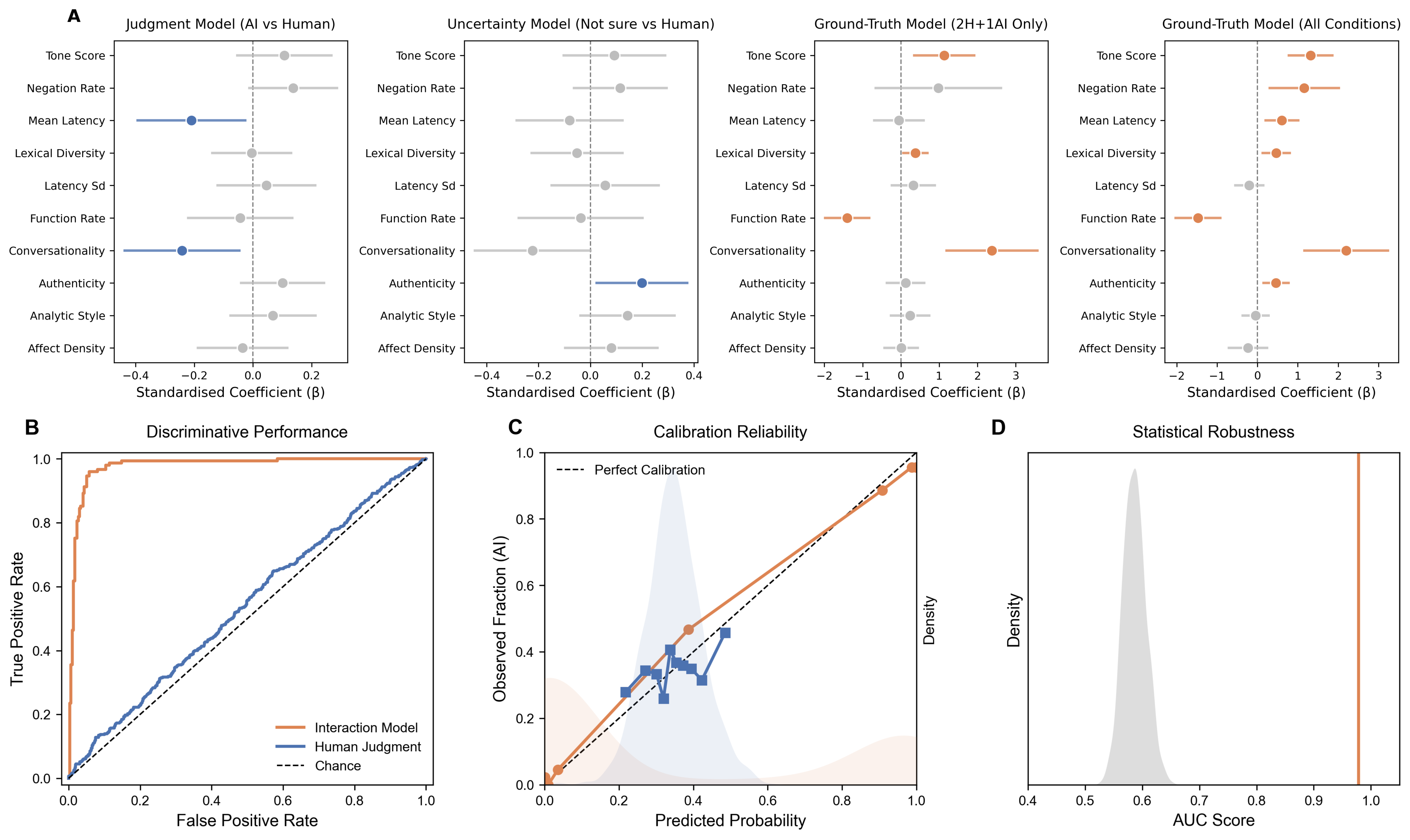}
    \caption{\textbf{Judgment-truth dissociation in cue-based identity inference.}
    \textbf{(A)} Forest plots show standardised coefficients ($\beta$) with 95\% confidence intervals from four logistic models: participants’ AI vs.\ Human judgments, uncertainty judgments (``Not sure'' vs.\ Human), ground-truth identity in mixed-identity triads (2H+1AI only), and ground-truth identity across all conditions. Vertical dashed lines indicate $\beta=0$; coloured points denote coefficients whose confidence intervals exclude zero.
    \textbf{(B)} Receiver operating characteristic (ROC) curves comparing human identity judgments (blue) with the dialogue-based interaction model predicting ground-truth identity (orange). The dashed diagonal denotes chance performance (AUC = 0.50).
    \textbf{(C)} Reliability diagram plotting observed fractions of AI targets against predicted probabilities (decile-binned). The interaction model closely follows the ideal calibration line (dashed), whereas human judgments show systematic miscalibration. Shaded densities indicate the distribution of predicted probabilities.
    \textbf{(D)} Null distribution of AUC scores from permutation tests preserving within-group structure ($N=1{,}000$). The vertical orange line marks the observed AUC of the interaction model, which lies far in the upper tail of the null distribution.}
    \label{fig:prediction}
\end{figure}

Across judgments (AI vs human; excluding ``not sure''), cue-based prediction was statistically detectable but weak in magnitude (Fig.~\ref{fig:prediction}A). In the logistic model of AI-attribution ($N=1{,}282$), longer mean response latencies were associated with a lower likelihood of judging a teammate as AI ($\beta=-0.210$, SE$=0.094$, $z=-2.25$, $p=.025$, 95\% CI $[-0.394,-0.027]$), while greater conversationality also decreased AI-attribution ($\beta=-0.242$, SE$=0.100$, $z=-2.42$, $p=.015$, 95\% CI $[-0.438,-0.046]$). Negation rate showed a trend towards increased AI-attribution ($\beta=0.137$, SE$=0.076$, $z=1.80$, $p=.072$, 95\% CI $[-0.012,0.286]$), whereas authenticity, function-word rate, lexical diversity, affect density, tone score, latency variability, and analytic style were not credibly associated with AI-attribution given the uncertainty in their CIs (all $p>.10$). Although omnibus model fit was non-zero (likelihood ratio test $p=.023$), explanatory power was small (McFadden’s pseudo-$R^2=0.018$), indicating that participants’ identity judgments only weakly tracked these dialogue cues once task and condition were accounted for, motivating a direct comparison with cue diagnosticity for ground truth.

Uncertainty judgments (``not sure'' vs.\ human; excluding AI-attributions) showed a different cue profile that reflected participants’ explicit reporting of indeterminacy rather than graded categorical classification. In this study, ``not sure'' responses were offered as a distinct response option and were used by participants to signal an inability to commit to either identity category, allowing uncertainty to be analysed separately from directional misclassification. In the logistic model contrasting ``not sure'' with human judgments ($N=1{,}072$), higher perceived authenticity of the target’s dialogue was associated with greater uncertainty ($\beta=0.199$, SE$=0.089$, $z=2.24$, $p=.025$, 95\% CI $[0.025,0.374]$), and conversationality showed a marginal association with reduced uncertainty ($\beta=-0.223$, SE$=0.114$, $z=-1.96$, $p=.050$, 95\% CI $[-0.445,0.000]$). Other cues were not credibly different from zero (all $p>.10$). The overall model did not improve substantially over the null (likelihood ratio test $p=.361$, pseudo-$R^2=0.015$), suggesting that uncertainty was only minimally structured by the aggregated cue set, consistent with participants reporting indeterminacy even when interaction appears ``natural'' by surface heuristics.

To quantify cue diagnosticity independent of participants’ beliefs, we modelled ground-truth identity using the same standardised predictors, focusing on the mixed-identity design where AI and human targets co-occurred under identical structural conditions. Because group condition quasi-deterministically encodes identity in the one-human (H1) and all-human (H3) settings, we restricted the primary truth model to H2 groups (two humans and one AI) to avoid structural separation; this yields a stringent test of whether cues discriminate AI from human teammates when the interaction context is held constant. In this H2-only truth model ($N=689$ judgments; 199 groups), several cues robustly distinguished AI from human targets: lower function-word rate strongly predicted AI identity ($\beta=-1.401$, SE$=0.294$, $z=-4.76$, $p<.001$, 95\% CI $[-1.978,-0.824]$), higher lexical diversity predicted AI identity ($\beta=0.380$, SE$=0.160$, $z=2.38$, $p=.017$, 95\% CI $[0.067,0.693]$), higher tone score predicted AI identity ($\beta=1.134$, SE$=0.400$, $z=2.84$, $p=.005$, 95\% CI $[0.350,1.917]$), and higher conversationality showed a large association with AI identity ($\beta=2.380$, SE$=0.607$, $z=3.92$, $p<.001$, 95\% CI $[1.190,3.570]$). Other cues (authenticity, affect density, latency metrics, negation, analytic style) were not reliably diagnostic in this constrained setting (all $p>.10$). Crucially, this truth model achieved substantial discrimination (pseudo-$R^2=0.692$), demonstrating that interactional cues contained separable identity information even when participants’ judgments did not track it strongly.

A complementary truth model estimated on the full sample without conditioning variables (and without group-condition adjustment) yielded a highly similar pattern and confirmed consistent cue directions under maximal power. In the full-sample truth model ($N=1{,}517$), authenticity ($\beta=0.461$, 95\% CI $[0.150,0.772]$, $p=.004$), lexical diversity ($\beta=0.463$, 95\% CI $[0.128,0.798]$, $p=.007$), tone ($\beta=1.320$, 95\% CI $[0.781,1.859]$, $p<.001$), mean latency ($\beta=0.605$, 95\% CI $[0.201,1.010]$, $p=.003$), negation ($\beta=1.159$, 95\% CI $[0.302,2.017]$, $p=.008$), and conversationality ($\beta=2.199$, 95\% CI $[1.167,3.230]$, $p<.001$) predicted AI identity, while function-word rate remained strongly negative ($\beta=-1.474$, 95\% CI $[-2.030,-0.919]$, $p<.001$), with pseudo-$R^2=0.716$. Because this full-sample specification aggregated across structurally different conditions, we treated it as a robustness check rather than the primary diagnosticity estimate; nevertheless, it converged with the constrained H2 test in showing that multiple social-heuristic cues differentiated AI from human dialogue at scale. The key implication is that the same cues that were statistically informative about identity in the corpus were only weakly incorporated into participants’ attributions, linking the results back to a central computer-mediated communication problem: identity inferences relied on heuristics that were only imperfectly mapped onto diagnostic features available in text-based interaction.

\subsection*{Robustness of judgment-truth dissociation}

Robustness analyses revealed a pronounced dissociation between behavioural diagnosticity, defined here as the extent to which interactional cues in dialogue reliably encode ground-truth identity, and participants’ explicit identity judgments. Focusing on \emph{complete-evaluation H2 triads} (2 humans + 1 AI; $N = 149$ triads, 447 evaluated teammates), target-level models using dialogue-derived interactional cues identified AI teammates with near-ceiling accuracy under group-wise cross-validation (mean AUC $= 0.983$, 95\% CI [$0.954$, $0.999$]; Brier score $= 0.046$, 95\% CI [$0.033$, $0.079$]; Fig.~\ref{fig:prediction}B). Cross-validated calibration indicated reasonable probability alignment, with a calibration slope of $0.75$ (95\% CI [$0.61$, $0.89$]), intercept of $-0.08$ (95\% CI [$-0.21$, $0.05$]), and low expected calibration error (ECE $= 0.028$; Fig.~\ref{fig:prediction}C). As a sensitivity check, we additionally examined calibration under a full-sample fit of the same target-level truth model. As expected for in-sample estimation, full-sample performance was slightly optimistic (AUC $= 0.988$, Brier score $= 0.033$), but calibration closely approached the ideal (slope $= 1.00$, intercept $= 0.001$, ECE $= 0.019$), indicating that the cross-validated reliability pattern was not driven by fold-specific artefacts. This diagnosticity was not attributable to response latency. Excluding latency-based features yielded virtually identical discrimination under group-wise cross-validation (AUC $= 0.983$, 95\% CI [$0.955$, $0.999$]; $\Delta$AUC $= -0.001$, 95\% CI [$-0.006$, $0.004$]), demonstrating that non-temporal linguistic and interactional cues alone were sufficient to distinguish AI from human teammates. Permutation tests preserving the H2 group structure produced a null AUC distribution centred substantially below the observed value (null mean $= 0.585$, 95\% CI [$0.547$, $0.629$]), with the empirical AUC lying far in the upper tail (permutation $p<.001$; Fig.~\ref{fig:prediction}D).

At the group level, a Top-1 identification analysis, selecting the single teammate with the highest predicted probability per triad, correctly identified the AI teammate in $95.3\%$ of triads (95\% CI [$0.91$, $1.00$]), far exceeding the chance baseline of $33.3\%$. Performance remained comparably high when timing features were excluded ($96.6\%$, 95\% CI [$0.92$, $1.00$]), confirming robustness across model specifications. In contrast, judgment-level models predicting participants’ explicit identity judgments showed no meaningful discriminative power. Models distinguishing ``AI'' from ``human'' judgments performed near chance (AUC $= 0.527$, 95\% CI [$0.48$, $0.57$], as did models predicting ``not sure'' responses (AUC $= 0.500$, 95\% CI [$0.45$, $0.55$]). Calibration for judgment models was poor, with shallow slopes and higher ECE values, indicating that predicted probabilities were weakly related to observed judgments. Together, these results demonstrate that while interactional cues reliably encode AI identity at the structural level, human observers largely fail to exploit this information when forming explicit identity judgments.

\subsection*{Representational dissociation}

To explain how identity discrimination can fail despite the presence of diagnostic interactional cues, we conducted a representational similarity analysis (RSA) comparing target-level spaces derived from interactional dialogue cues, modal participant identity judgments, ground-truth identity, subjective impressions (humanness and trust), and the semantic structure of impression narratives. RSA evaluates whether pairs of targets that are similar in one space (e.g., dialogue features) are also similar in another (e.g., identity labels), thereby characterising the geometry of social inference beyond accuracy-based metrics.

\begin{figure}
    \centering
    \includegraphics[width=1\linewidth]{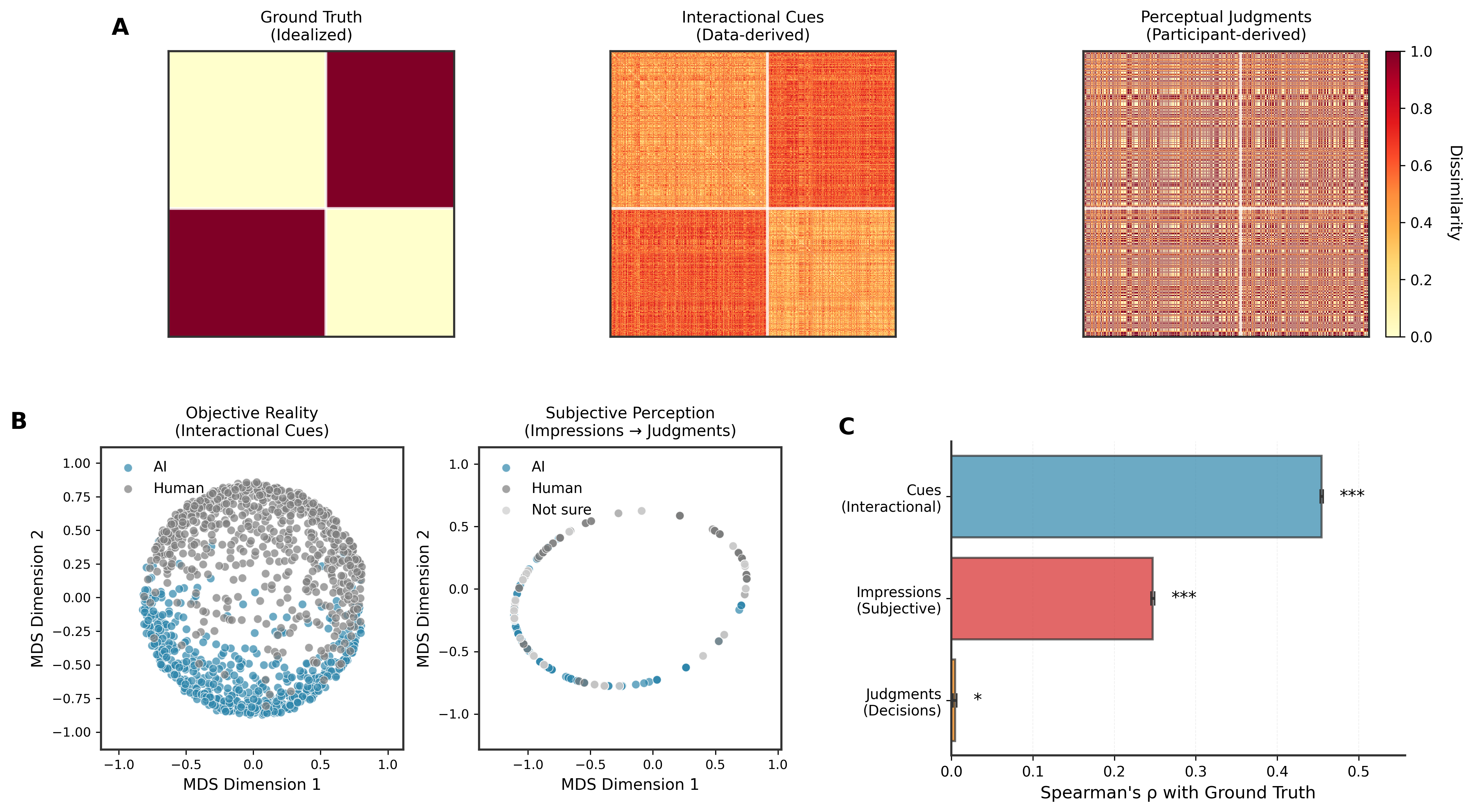}
    \caption{\textbf{Representational dissociation between objective interactional structure and subjective identity inference.}
            \textbf{(A)} Representational dissimilarity matrices (RDMs) for ground-truth identity (idealised), interactional cues (data-derived), and perceptual judgments (participant-derived). 
            \textbf{(B)} Two-dimensional multidimensional scaling (MDS) embeddings of targets derived from interactional cue space (left) and perceptual judgment space (right). 
            \textbf{(C)} Spearman correlations between each representational space and ground-truth identity. Error bars denote 95\% bootstrap confidence intervals; asterisks indicate permutation-based significance ($^{*}p<.05$, $^{***}p<.001$).
            }
    \label{fig:rsa}
\end{figure}

Across 1{,}148 targets with complete data, interactional cue structure was strongly aligned with ground-truth identity (Spearman $\rho = .455$, 95\% bootstrap CI [.453, .456], permutation $p<.001$), indicating that AI and human teammates occupied systematically separable regions in cue space (Fig.~\ref{fig:rsa}A). In contrast, participants’ identity judgments bore no relation to this interactional structure ($\rho \approx 0$, 95\% CI [$-.002$, .002], permutation $p= .924$) and showed only a negligible alignment with ground truth ($\rho = .004$, 95\% CI [.002, .007], permutation $p= .027$). Instead, judgment structure aligned with subjective impressions: the representational space defined by humanness and trust ratings showed a small but reliable correspondence with judgments ($\rho = .247$, 95\% CI [.245, .250], permutation $p<.001$), whereas impression structure itself was effectively orthogonal to interactional cues ($\rho = .007$, 95\% CI [.004, .009]). Two-dimensional multidimensional scaling (MDS) embeddings illustrated this dissociation: AI and human targets separated clearly in interactional cue space but were intermingled in perceptual judgment space, which was organised primarily by subjective impressions rather than by actual identity (Fig.~\ref{fig:rsa}B).

The dissociation between interactional cues that reliably indexed ground-truth identity and participants’ identity judgments was not resolved by participants’ explicit justifications. A topic-based representational space derived from BERTopic assignments showed only negligible correspondence with judgment structure ($\rho = .025$, 95\% CI [.022, .027]) and no correspondence with cue structure ($\rho \approx 0$), indicating that the semantic themes used to explain identity impressions were weakly related to categorical attributions and entirely decoupled from diagnostic interactional features. Crucially, the same pattern persisted under a stringent test restricted to mixed-identity triads (2H+1AI), where AI and human targets interacted under identical structural conditions: cue structure remained strongly aligned with ground truth ($\rho = .415$, 95\% CI [.410, .419]), whereas judgment structure remained only negligibly related to cues ($\rho = .008$, 95\% CI [.003, .013]) and to ground truth ($\rho = .031$, 95\% CI [.026, .037]). This contrast is summarised quantitatively in Fig.~\ref{fig:rsa}C. Together, these results provide a geometric account of the judgment-truth dissociation observed throughout the results: identity-relevant information is robustly structured in interactional behaviour, but participants organise social inference in a distinct representational space anchored in subjective impressions rather than in the diagnostic cues available in synchronous text-based collaboration.

\section*{Discussion}

Across synchronous, text-based group interactions, participants did not discriminate undisclosed AI teammates from human teammates above chance, despite interacting long enough to form clear social impressions. This collapse of identity discrimination was consistent across task domains and persisted in the cleanest within-group comparison setting, mixed-identity triads (2H+1AI), where AI and human targets co-occurred under identical structural conditions. Importantly, the failure of identity judgment did not reflect an absence of social sensitivity: participants reliably differentiated conversational stances and calibrated perceived trust and humanness accordingly. Instead, the results converged on a systematic judgment-truth dissociation. Interactional cues in the dialogue reliably encoded ground-truth identity, allowing AI and human teammates to be distinguished with high accuracy under stringent within-group validation. However, participants’ explicit identity judgments incorporated these cues only weakly and inconsistently, revealing a systematic dissociation between behavioural diagnosticity and social judgment. At the representational level, identity judgments were geometrically decoupled from interactional cue structure and organised primarily around subjective impressions. Together, these findings indicate that, under truth-default conditions, the heuristics people use to infer identity in computer-mediated communication are no longer reliably aligned with the behavioural regularities that encode identity in contemporary online interaction.

Classic accounts of computer-mediated communication emphasise that humans adapt to cue-lean environments by treating available textual and temporal signals as substitutes for filtered nonverbal information, enabling interpersonal inference and relational development despite bandwidth constraints \cite{walther1992social,walther2002cues}. Our results extend this perspective by identifying a boundary condition: the same adaptive reliance on linguistic and chronemic heuristics can fail to support categorical identity inference when sophisticated conversational agents are present as undisclosed social actors. In our data, participants’ post-interaction explanations and supervised linguistic contrasts suggest heavy reliance on surface cues such as response latency, perceived fluency, and generic humanness language, signals that have historically served as plausible indicators of social presence in computer-mediated communication \cite{walther2002cues,hancock2020ai}. Yet these heuristic cues were weakly aligned with ground truth once task and experimental context were accounted for, yielding near-chance discrimination even as participants remained socially responsive to stance and interpersonal tone. This pattern supports a refinement of social information processing accounts: the capacity to form coherent social impressions from limited-bandwidth interaction can remain intact while the mapping from those impressions to source identity becomes unreliable \cite{wangSurveyLargeLanguage2024a,asheryEmergentSocialConventions2025}.

Truth-default theory posits that in ordinary interaction people default to presuming honesty and benign intent unless contextual triggers raise suspicion \cite{levine2014truth,levine2022truth}. The present findings show that a comparable truth-default operates for presumed human identity in synchronous text-based interaction. Participants exhibited a slight bias toward human attribution and expressed uncertainty frequently, yet neither uncertainty nor task demands reliably improved discrimination sensitivity. The confusion structure was also informative: socially fluent AI agents were often accepted as human, while competent humans were frequently misclassified as AI. This symmetry is difficult to reconcile with accounts that treat AI detection as a simple function of noticing ``machine-like'' errors. Instead, it suggests that under truth-default conditions participants do not systematically translate interactional anomalies into categorical identity conclusions, and that the cues they treat as suspicious (for example, perceived scriptedness or rapid response) can be shared by both AI and human targets. In other words, the social inference problem is not simply that AI is increasingly fluent; it is that the evidential value of commonly used suspicion cues has become unstable in the contemporary ecology of online interaction \cite{hancock2020ai,jakesch2023human,williamsLargeLanguageModels2025}.

A central contribution of this work is to distinguish \textit{availability} of identity-relevant information from \textit{use} of that information in human judgment. Cue-based models predicting participants’ explicit identity judgments performed near chance, with small effect sizes and limited explanatory power, implying that participants’ judgments only weakly tracked aggregated interactional cues after adjusting for task and condition. In sharp contrast, models predicting ground-truth identity from the same cue set achieved near-ceiling discrimination in mixed-identity triads, including under group-wise cross-validation and under specifications excluding latency-based predictors. These results establish that identity information is robustly structured in interactional behaviour and not reducible to simple timing artefacts. The dissociation therefore reflects a mismatch between the interactional features that encode identity and the heuristics participants deploy when making explicit identity judgments. This mismatch is consistent with prior evidence that human heuristics for detecting AI-generated language are systematically flawed \cite{jakesch2023human}. Our study extends that conclusion to synchronous group interaction, showing that even when identity is strongly recoverable from behavioural traces at the target level, human observers do not reliably extract or integrate that information into explicit identity judgments.

The representational similarity analysis provides a structural explanation for how identity inference can fail despite the presence of diagnostic cues (Fig.~\ref{fig:rsa}). Interactional cue space was strongly aligned with ground truth, indicating that AI and human targets occupied separable regions in behavioural space. Yet participants’ judgment space was effectively orthogonal to this interactional structure and showed negligible correspondence with ground truth. Instead, judgment structure aligned with subjective impressions derived from trust and humanness ratings. This pattern suggests that, in synchronous text-based interaction, identity judgments are not organised around the behavioural dimensions that best separate AI from humans; they are organised around a perceptual space anchored in evaluative impressions. Such an account reconciles the coexistence of (i) robust stance sensitivity and socially meaningful evaluations and (ii) chance-level categorical identity discrimination. It also helps interpret the limited diagnostic value of participants’ explicit justifications: semantic themes in impression narratives were only weakly related to judgment outcomes and were decoupled from interactional cue structure. In sum, identity inference appears to be governed by a representational geometry that privileges interpersonal evaluation over diagnostic discrimination. This structural account complements accuracy-based analyses by specifying \textit{where} in the inference process the decoupling emerges: not at the level of social perception per se, but at the mapping from perceptual organisation to ground-truth identity.

The findings also speak to established accounts of anthropomorphism in human-machine interaction. The CASA paradigm predicts that humans apply social rules to interactive media when cues support social responding \cite{nass2000machines}. Contemporary work similarly suggests that agentic, socially fluent systems can elicit parasocial engagement and anthropomorphic interpretations \cite{maeda2024human}. Consistent with these accounts, conversational cues in our study reliably evoked social evaluations: participants clearly differentiated supportive versus contrarian stances and calibrated perceived trust and humanness accordingly. However, the results also clarify an important boundary of anthropomorphic responding, namely that social responsiveness does not entail accurate categorical identity inference. Under undisclosed conditions, participants may treat an interlocutor as socially present while remaining unable to ground that impression in a correct judgment of source identity. This distinction matters for interpreting responses to AI-mediated communication: measures of trust, humanness, or engagement can vary meaningfully with interactional style even when participants cannot (and do not) reliably infer whether the source is human or artificial \cite{hancock2020ai,go2019humanizing}. Related work on AI-mediated communication and human-AI collaboration has documented benefits for conversational quality and perceived empathy in supportive contexts \cite{sharma2022human,costelloDurablyReducingConspiracy2024}. Our results add that such socially consequential responses can occur in parallel with systematic uncertainty or misclassification about identity when disclosure is absent.

Identity inference is not merely a perceptual curiosity; it is foundational to how trust and accountability are distributed in online interaction. When interactional cues no longer reliably indicate whether an interlocutor is human, routine social judgments that rely on presumed identity can become epistemically fragile. This has direct relevance for digital environments in which conversational agents increasingly participate without explicit disclosure \cite{hancock2020ai,unesco2025freedom}. In political communication, for example, conversational agents can be used to test and deploy persuasive strategies at scale \cite{argyle2025testing,salvi2025conversational,wackGenerativePropagandaEvidence2025}. In organisational settings, emerging guidance suggests that identity verification and authentication may become unreliable in isolation when confronted with AI-generated content and deepfakes \cite{gartner2024deepfakes, bengioManagingExtremeAI2024}. These contexts underscore a broader shift toward ``machine behaviour'' as a domain in which AI agents act within human networks and shape collective outcomes \cite{rahwan2019machine, brinkmann2023machine, barberaCriticalPeripheryGrowth2015, centolaExperimentalEvidenceTipping2018, asheryEmergentSocialConventions2025}. The present findings indicate that, under plausible interaction conditions, human observers may be systematically unable to distinguish artificial from human social actors even when interactional traces contain diagnostic information. This mismatch complicates informal accountability mechanisms that assume interlocutor identity can be inferred from conversational behaviour and highlights the epistemic stakes of disclosure in AI-mediated communication.

The findings are bounded to undisclosed, synchronous, text-based group interaction in a custom computer-mediated communication environment. First, identity inference may differ when richer modalities are available (e.g., voice, video, or persistent identity signals), or when participants are explicitly warned that AI may be present, which would alter truth-default conditions \cite{levine2014truth,levine2022truth}. Second, the AI agents were engineered to be socially fluent and to follow stance-specific protocols; different agent designs or task structures may yield different cue profiles and thus different degrees of misclassification. Third, our analyses focus on immediate post-interaction judgments rather than learning or adaptation across repeated exposures. Work on repeated interaction with large language models suggests that strategic dynamics can evolve across time, implying that experience may shape how people interpret machine behaviour in interactive settings \cite{akata2025playing}. 

\section*{Materials and Methods}

\subsection*{Participants}

Participants were recruited through Prolific and were required to be 18--65 years old and fluent in English. This recruitment strategy was intended to yield a heterogeneous adult sample familiar with online text-based interaction, consistent with prior work on computer-mediated communication and collaborative problem solving. Prolific safeguards were used to prevent duplicate participation, and participants received financial compensation consistent with ethical guidance for online behavioural research. Ethical approval was obtained from Monash University (Project ID: 48379). All participants provided informed consent before participation and were fully debriefed after the study regarding the presence of AI teammates.

A total of 905 participants were enrolled. After excluding incomplete sessions, failed group matching, and records without usable interaction data, the final analytic sample comprised 786 unique participants nested within 471 collaborative groups. These participants generated 1,572 teammate-level identity judgments following the collaborative phase. Of these judgments, 685 targeted AI teammates and 887 targeted human teammates. No identity-judgment responses were missing, and 257 judgments (16.3\%) were recorded as \textit{Not sure}, allowing uncertainty to be analysed separately from directional classification. Full demographic characteristics of the recruited and analysed samples are reported in the Supplementary Materials.

\subsection*{Experimental design}

The study used a mixed factorial design implemented in a custom-built web platform for real-time, synchronous, text-based collaboration. The design crossed \textit{task domain} (analytical, ethical, creative) with \textit{group composition / AI stance condition}, yielding an effective \(3 \times 5\) structure (Fig.~\ref{fig:method}). The goal was to test whether participants could infer teammate identity under conditions typical of contemporary online communication, where interaction is text-only and no physical or vocal cues are available.

Groups were assigned to one of five conditions. In the human-only control condition (3H), triads consisted entirely of human participants. In mixed groups, triads contained either two humans and one AI agent (2H+1AI) or one human and two AI agents (1H+2AI). When present, AI agents adopted one of two pre-specified conversational stances: \textit{supportive} or \textit{contrarian}. Supportive agents provided affirmation, encouragement, and consensus-oriented scaffolding, whereas contrarian agents introduced scepticism, critique, and counterarguments. In 1H+2AI groups, both AI agents shared the same stance. This design enabled us to test identity inference across different levels of AI exposure and across qualitatively different interaction styles.

\subsection*{Collaborative tasks}

Participants completed one of three collaborative tasks selected to elicit distinct forms of group cognition. The \textit{analytical} task was an adapted winter survival-ranking exercise requiring convergent reasoning against an expert benchmark \cite{Johnson1987,Rogelberg1992}. The \textit{ethical} task was an autonomous-vehicle moral dilemma requiring value-based reasoning among competing outcomes \cite{Bonnefon2016,Bigman2020Nature,Awad2018}. The \textit{creative} task was a collaborative movie-plot ideation activity requiring divergent idea generation around a prompt in which artificial general intelligence played a central narrative role \cite{Korde2017,Hwang2021IdeaBot,Farrokhnia2025}. Together, these tasks sampled analytical convergence, normative deliberation, and creative ideation. All tasks were delivered through the same interface and under the same time constraints to ensure comparability across domains. Full task materials and benchmark information are provided in the Supplementary Materials.

\subsection*{Procedure and masking protocol}

Each session comprised three stages: an initial individual familiarisation phase, a synchronous group discussion, and a post-collaboration assessment. Participants first reviewed and began the assigned task individually so that all group members entered the discussion with the same problem context. They then engaged in a 10-minute real-time text discussion with their assigned teammates. Following the collaborative phase, participants completed post-interaction ratings and identity judgments for each teammate.

To preserve identity ambiguity, all human participants and AI agents were assigned neutral pseudonyms (for example, Kevin, Stuart, and Bob). Participants were told that they would collaborate online with other participants, but they were not informed that some teammates could be AI agents. This non-disclosure was central to the inferential setting of interest: participants had to rely on interactional traces alone when judging whether a teammate was human or artificial. Immediately after the discussion, participants rated each teammate’s humanness, trustworthiness, supportiveness, and conflictuality, then indicated whether they believed each teammate was \textit{Human}, \textit{AI}, or \textit{Not sure}, and finally provided a brief open-ended explanation for that judgment. At the end of the study, participants were fully debriefed about the experimental use of AI teammates. No adverse reactions were reported.

\subsection*{AI agent implementation}

AI agents were designed to function as plausible ordinary teammates in synchronous text-based collaboration while systematically varying conversational stance. Each agent was instantiated as either \textit{supportive} or \textit{contrarian}, based on prior work showing that interactional stance shapes perceived trust, coordination, conflict, and group dynamics in computer-mediated interaction \cite{Hwang2021IdeaBot,Edmondson1999,walther1992social,walther2002cues}. Supportive agents used affirming, inviting, and consensus-building language; contrarian agents used sceptical, challenging, and dissenting language. Across conditions, agents were instructed never to reveal or imply that they were artificial.

Agent behaviour was implemented through structured system prompts specifying persona, tone, response length, and character-maintenance rules. To enhance ecological validity, prompts instructed agents to use casual conversational language, occasional imperfections, contractions, uncertainty, and self-correction, thereby reducing the stylised regularity often associated with machine-generated dialogue. The prompts also constrained the agents to maintain a human-like stance if questioned directly about identity. We did not use task-specific fine-tuning or bespoke deception-optimised architectures; instead, the design intentionally relied on prompt-based behavioural control to test whether socially fluent undisclosed participation could emerge from comparatively accessible agentic scaffolding. Full system prompts are provided in the Supplementary Materials.

To approximate natural participation timing, agent contributions were governed by a probabilistic scheduling protocol rather than deterministic turn-taking. Each agent scanned the group chat at a base interval of 25 seconds with a uniformly random jitter of up to \(\pm 25\%\) applied to each scan. At each scan, the agent generated a message with probability \(p = 0.5\). If two AI agents attempted to respond simultaneously in the same group, one was randomly selected to respond and the other was delayed by 10 seconds to avoid unrealistic simultaneity. To prevent artificial dominance, agents were limited to a maximum of three consecutive messages without an intervening contribution from another group member. These rules were intended to reproduce the bounded irregularity of human participation while maintaining comparable participation rates across groups.

The naturalness of this protocol was evaluated in an independent pilot study (\(N = 15\)) using the same task environment. After interacting with the agents, pilot participants rated partner humanness on a 7-point scale. Agents received relatively high humanness ratings (\(M = 5.43\), \(SD = 1.07\)), suggesting that the combination of prompt-based behavioural control and probabilistic timing produced interaction patterns that participants experienced as plausible. Additional implementation details are provided in the Supplementary Materials.

\subsection*{Post-collaboration evaluations}

For each teammate, participants completed four social evaluations: perceived humanness, trust, supportiveness, and conflictuality. Supportiveness and conflictuality were measured on 7-point Likert scales (\(1 =\) strongly disagree, \(7 =\) strongly agree); humanness and trust were measured using corresponding Likert-type ratings. These dimensions were selected because they capture core components of social evaluation in text-based collaboration, including perceived social presence, interpersonal trust, and interactional stance.

Participants then made a categorical identity judgment for each teammate using three response options: \textit{Human}, \textit{AI}, or \textit{Not sure}. Finally, they provided a short free-response explanation describing the cues or impressions that informed their judgment. These open-ended texts were used to characterise the semantic heuristics participants reported relying on when inferring identity.

\subsection*{Dialogue-derived interactional cues}

All chat exchanges were logged at the utterance level with timestamps, speaker identifiers, message text, and word counts. To quantify the behavioural traces available for identity inference, we derived a set of interactional cues at the group--target level, where each target corresponded to one teammate within one collaborative group. This level of aggregation was chosen because participants judged teammates, not isolated utterances, and because teammate identity was expressed through an accumulated interaction profile across the discussion.

We extracted three classes of features: dictionary-based linguistic cues, temporal responsiveness, and lexical diversity. Linguistic and stylistic cues were based on LIWC-style variables available at the utterance level, specifically \textit{Authentic}, \textit{function}, \textit{Affect}, \textit{Tone}, \textit{negate}, \textit{Analytic}, and \textit{Conversation}. For each target, these variables were aggregated using word-count-weighted means so that longer utterances contributed proportionally more to the final estimate. Temporal responsiveness was indexed by response latency between consecutive messages; for each target we computed both mean latency and latency variability, excluding latencies that were undefined by construction, such as the first message in a thread. Lexical diversity was computed on each target’s concatenated message stream in timestamp order using the Measure of Textual Lexical Diversity. We also retained message count and total word count as exposure controls.

The resulting target-level cue profiles were merged into the judgment-level dataset by matching group identifier and teammate identity, so that each participant judgment was paired with the corresponding behavioural profile of the judged teammate. Cue variables were standardised before modelling. Detailed preprocessing rules, variable definitions, aggregation formulas, and missing-data handling are reported in the Supplementary Materials.

\subsection*{Signal detection and individual-difference analyses}

To quantify participants’ ability to distinguish AI from human teammates, we used signal detection theory (SDT), treating AI targets as signal and human targets as noise. Identity responses were collected using three categories (\textit{AI}, \textit{Human}, \textit{Not sure}). For SDT analyses, AI judgments were treated as positive responses and Human judgments as negative responses; \textit{Not sure} responses were retained for descriptive reporting but were not forced into either category.

Sensitivity was quantified using
\[
d'=\Phi^{-1}(H^{*})-\Phi^{-1}(F^{*}),
\]
where \(H^{*}\) is the corrected hit rate for judging an AI target as AI and \(F^{*}\) is the corrected false-alarm rate for judging a human target as AI. To avoid infinite \(z\)-scores, log-linear corrections were applied when hit or false-alarm rates equalled 0 or 1:
\[
H^{*}=\frac{\mathrm{hits}+0.5}{N_{\mathrm{AI}}+1}, \qquad
F^{*}=\frac{\mathrm{false\ alarms}+0.5}{N_{\mathrm{H}}+1}.
\]
Response bias was indexed by
\[
\beta=\exp\left(-z_H d' + 0.5d'^2\right),
\]
where \(z_H=\Phi^{-1}(H^{*})\).

We estimated SDT indices for the full sample and stratified them by task domain and group condition where both AI and human targets were present. Hit-rate and false-alarm confidence intervals were computed using Wilson score intervals, and confidence intervals for \(d'\) were estimated via nonparametric bootstrap resampling (1,000 iterations). To test whether aggregate near-chance performance masked better-performing individuals, we also computed participant-level \(d'\) values from each participant’s own set of judgments and summarised their distribution across the sample.

In addition, we examined whether sensitivity was associated with individual differences. Participant-level \(d'\) values were correlated with AI literacy and Big Five personality measures using Pearson correlations with confidence intervals based on Fisher’s \(z\) transformation. Demographic differences in \(d'\) were examined using independent-samples tests or one-way analyses of variance as appropriate. Finally, to validate the stance manipulation, we compared supportiveness and conflictuality ratings between supportive and contrarian AI targets and also compared humanness, trust, supportiveness, and conflictuality between AI and human targets. Full descriptive statistics, subgroup outputs, and effect sizes are reported in the Supplementary Materials.

\subsection*{Computational analysis of impression texts}

Participants’ open-ended impression texts were analysed to identify the semantic heuristics used when justifying identity judgments. Empty or whitespace-only responses were excluded. For the retained texts, we first computed descriptive statistics including character length and word count, and assessed whether excluding empty texts materially altered the distribution of judgment categories.

We then used BERTopic to identify recurrent semantic themes in the impression corpus. Texts were embedded using the pretrained sentence-transformer model \textit{all-MiniLM-L6-v2}, reduced using UMAP, and clustered using HDBSCAN. Topic representations were generated with class-based TF--IDF to produce interpretable keyword summaries. Documents assigned to topic \(-1\) were treated as outliers and excluded from inferential topic analyses. Topic prevalence was estimated with 95\% confidence intervals.

To assess whether these semantic themes mapped onto judgment outcomes, we analysed the association between topic membership and judgment category using contingency analyses and quantified effect sizes with Cramér’s \(V\), with bootstrap confidence intervals. We also estimated mutual information between topic assignments and judgment categories as an overall index of how informative topic structure was for explaining judgments. To complement the unsupervised topic model, we conducted supervised class-based TF--IDF analyses to identify lexically distinctive terms associated with different judgment categories. Topic-level odds ratios comparing incorrect and correct judgments and multinomial models predicting judgment category from topic membership were used as additional descriptive checks. Full hyperparameters, topic diagnostics, similarity analyses, and complete statistical outputs are provided in the Supplementary Materials.

\subsection*{Prediction of identity judgments and ground-truth identity}

To test whether participants’ explicit identity judgments tracked the behavioural cues available in the dialogue, and to contrast this with the actual diagnosticity of those same cues for ground-truth identity, we estimated a set of logistic regression models using the dialogue-derived predictors described above. All cue variables were \(z\)-standardised before modelling.

We first estimated two judgment-level models. The \textit{AI-attribution} model was restricted to judgments labelled \textit{AI} or \textit{Human}, with AI coded as 1 and Human as 0. The \textit{uncertainty} model was restricted to judgments labelled \textit{Not sure} or \textit{Human}, with \textit{Not sure} coded as 1 and Human as 0. Both models used the same predictor set and included task type and group condition as covariates, enabling direct comparison of cue profiles associated with directional attribution versus explicit uncertainty. Because each participant judged multiple teammates and judgments were nested within collaborative groups, primary inference used cluster-robust standard errors clustered at the participant level.

To quantify cue diagnosticity independently of participants’ beliefs, we also modelled ground-truth identity using the same predictors. The primary truth model was restricted to mixed-identity groups (2H+1AI), where AI and human targets co-occurred in the same structural setting; this avoided the quasi-deterministic encoding of identity by group composition in human-only or AI-majority conditions. In this restricted model, task type was retained as a covariate, whereas group condition was omitted because it was constant within the subset. Truth models used cluster-robust standard errors at the group level. Model fit was summarised using likelihood-based indices and pseudo-\(R^2\). Full coefficient tables and additional specifications are reported in the Supplementary Materials.

\subsection*{Robustness and sensitivity analyses}

We conducted a series of robustness analyses to test whether the observed dissociation between judgment and ground-truth prediction depended on clustering choices, sparse targets, timing features, or between-group confounding. For judgment models, we re-estimated standard errors using group-level rather than participant-level clustering and examined collinearity using variance inflation factors. We also repeated analyses after excluding low-information targets defined by the bottom decile of total word count.

For the truth models, we constructed a target-level dataset with one row per evaluated teammate and focused on a stringent subset of complete-evaluation mixed-identity triads, defined as mixed triads in which all three teammates received at least one human evaluation and exactly one teammate was AI. Within this universe, we estimated within-group diagnosticity using a group-stratified conditional logistic regression, with a fallback group fixed-effects logistic regression when needed for numerical stability. This specification identifies AI targets from within-group variation alone and is therefore invariant to between-group differences.

We then evaluated out-of-sample discrimination using group-wise cross-validation that held out entire groups per fold, thereby preventing information leakage across members of the same interaction context. Model performance was summarised using the area under the receiver operating characteristic curve, Brier score, calibration slope, calibration intercept, and expected calibration error. To assess whether diagnosticity was driven mainly by timing, we repeated the within-group and cross-validated analyses after removing latency-related predictors. We also performed a within-group permutation test that preserved the mixed-identity triad structure by randomly assigning exactly one AI label per triad, generating a null distribution for discrimination. Finally, for interpretability, we reported a triad-level Top-1 identification analysis in which the teammate with the highest predicted AI probability was selected as the model’s inferred AI target. Detailed outputs for all robustness checks are reported in the Supplementary Materials.

\subsection*{Representational similarity analysis}

To characterise how identity-relevant information was structured across behavioural traces, subjective judgments, and ground truth, we conducted a representational similarity analysis (RSA) at the target level. RSA tests whether pairs of targets that are similar in one representational space are also similar in another, thereby providing a geometric account of inference beyond conventional accuracy metrics.

Judgment-level observations were first aggregated to the target level, defined as a unique teammate within a collaborative group. For each target, we computed the modal identity judgment (\textit{AI}, \textit{Human}, or \textit{Not sure}), ground-truth identity, the mean dialogue-derived cue profile, mean subjective impression ratings (humanness and trust), and the modal semantic topic assigned from impression-text analysis. Targets with incomplete cue data were excluded from RSA.

We then constructed representational dissimilarity matrices for five spaces: interactional cue space, judgment space, ground-truth identity space, subjective impression space, and semantic topic space. Cue-based and impression-based matrices were computed as cosine distances over standardised feature vectors. Ground-truth and topic matrices used binary same/different coding, whereas judgment matrices used graded distances so that pairs involving \textit{Not sure} were treated as intermediate rather than maximally distinct. Alignment between spaces was quantified by correlating the upper triangles of the matrices using Spearman rank correlations. Statistical significance was assessed with permutation tests, and uncertainty was quantified with bootstrap confidence intervals obtained by resampling dissimilarity pairs. RSA was conducted both on the full target set and on the mixed-identity subset (2H+1AI), where AI and human targets shared the same interaction context. Multidimensional scaling plots were used for visualisation. Additional implementation details and robustness checks are provided in the Supplementary Materials.


\clearpage 

\bibliography{science_template} 
\bibliographystyle{sciencemag}


\section*{Acknowledgments}

\paragraph*{Funding:}
This research was supported by the National Natural Science Foundation of China (Grant No.~20261710003; L.Y.), the Australian Research Council (DP240100069 and DP220101209; D.G.), and the Jacobs Foundation (CELLA 2 CERES; D.G.).
\paragraph*{Author contributions:}
L.Y. conceived the study, designed the research, conducted the analyses, and drafted the manuscript. Y.J. contributed to data analysis and manuscript drafting. X.H. contributed to study design and data interpretation. D.G. contributed to conceptualisation, provided critical revisions, and supervised the research. All authors reviewed and approved the final manuscript.
\paragraph*{Competing interests:}
There are no competing interests to declare.
\paragraph*{Data and materials availability:}
Analysis code and scripts to reproduce the results and figures will be made publicly available after the acceptance of this paper.

\newpage


\renewcommand{\thefigure}{S\arabic{figure}}
\renewcommand{\thetable}{S\arabic{table}}
\renewcommand{\theequation}{S\arabic{equation}}
\renewcommand{\thepage}{S\arabic{page}}
\setcounter{figure}{0}
\setcounter{table}{0}
\setcounter{equation}{0}
\setcounter{page}{1} 


\begin{center}
\section*{Supplementary Materials for\\ \scititle}
Lixiang~Yan$^{\ast}$ \and 
Yueqiao~Jin \and
Xibin~Han \and 
Dragan~Gasevic \and
\small$^\ast$Corresponding author. Email: lixiangyan@tsinghua.edu.cn\and
\end{center}

\subsubsection*{This PDF file includes:}
Supplementary Methods\\
Supplementary Results\\
Figures S1 to S5\\
Tables S1 to S35\\

\newpage


\subsection*{Supplementary Methods}

This Supplementary section provides the procedural and analytic detail that supports the condensed Method reported in the main text. It includes four classes of information that are intentionally streamlined in the main article. First, it documents sample composition, demographic characteristics, and the full wording of the collaborative tasks used in the experiment. Second, it reports the full implementation logic of the AI agents, including persona prompts and the interaction protocol used to embed agents in synchronous group chat. Third, it specifies the preprocessing and feature-construction pipeline used to derive dialogue-based interactional cues from the raw utterance logs. Fourth, it provides the full technical details for all statistical and computational analyses, including signal detection computations, topic modelling of impression texts, logistic regression models for judgment and ground-truth prediction, robustness checks, and representational similarity analysis.

\subsubsection*{Demographics}

Table~\ref{tab:demographics} reports the demographic composition of the recruited sample. As noted in the main text, 905 participants were initially enrolled, of whom 786 were retained in the final analytic sample after exclusions due to incomplete participation, failed matching, or unusable interaction records. The recruited sample was intentionally heterogeneous with respect to gender, age, region, employment, and educational background, consistent with the study’s aim of approximating adult users of contemporary online communication environments rather than a narrowly defined convenience population.

The demographic profile also provides context for the individual-difference analyses reported in the main text. In particular, the broad spread across age groups, global regions, employment statuses, and educational attainment supports the interpretation that the observed near-chance identity discrimination was not restricted to a narrow subgroup of participants. Additional subgroup counts used in individual-difference analyses are reported in the Supplementary Results.

\begin{table}[ht]
\centering
\caption{Demographic characteristics of participants ($N = 905$).}
\label{tab:demographics}
\begin{tabular}{llll}
\toprule
Variable & Category & $n$ & \% \\
\midrule
Gender & Male        & 438 & 48.4 \\
       & Female      & 464 & 51.3 \\
       & Other       &   3 & 0.3 \\
Age    & 18--24      & 148 & 16.4 \\
       & 25--34      & 379 & 41.9 \\
       & 35--44      & 183 & 20.2 \\
       & 45--54      & 109 & 12.0 \\
       & 55--64      &  62 & 6.9 \\
       & 65+         &  24 & 2.7 \\
Region & North America/Central America & 135 & 14.9 \\
       & South America                 &  15 & 1.7 \\
       & Europe                        & 369 & 40.8 \\
       & Africa                        & 363 & 40.1 \\
       & Asia                          &  11 & 1.2 \\
       & Australia                     &  12 & 1.3 \\
Employment & Working full-time               & 629 & 69.5 \\
           & Working part-time               & 119 & 13.1 \\
           & Unemployed (seeking)            &  39 & 4.3 \\
           & Homemaker/parent                &  16 & 1.8 \\
           & Student                         &  61 & 6.7 \\
           & Retired                         &  20 & 2.2 \\
           & Other                           &  21 & 2.3 \\
Education  & High school                     & 130 & 14.4 \\
           & Vocational                      &  78 & 8.6 \\
           & Bachelor                        & 442 & 48.8 \\
           & Master                          & 217 & 24.0 \\
           & PhD                             &  38 & 4.2 \\
\bottomrule
\end{tabular}
\end{table}

\subsubsection*{Task materials}

Participants completed one of three collaborative tasks designed to elicit complementary forms of group cognition: convergent analytical reasoning, ethical deliberation, and divergent creative ideation. The rationale for including multiple task domains was not to compare task performance per se, but to test whether human identity inference generalised across qualitatively different collaborative settings. The tasks were selected because each has an established literature, produces active multi-party discussion, and can be implemented in a text-only online interface without requiring specialised prior knowledge.

The analytical task was an adapted winter survival exercise \cite{Johnson1987,Rogelberg1992}. It required participants to rank salvaged items according to their survival value under a constrained scenario with an expert benchmark. This task was chosen because it reliably elicits justificatory reasoning, disagreement, and convergence pressure. The ethical task was an autonomous-vehicle moral dilemma adapted from prior research on machine ethics and moral judgment \cite{Bonnefon2016,Bigman2020Nature,Awad2018}. It was chosen because it elicits normative reasoning without a uniquely correct answer and therefore creates space for stance-based variation in discussion. The creative task was a collaborative story-generation exercise adapted from group ideation paradigms \cite{Korde2017,Hwang2021IdeaBot,Farrokhnia2025}. It was included to test whether identity inference also fails in an open-ended, divergent setting where conversational naturalness and idea elaboration may matter more than factual or normative correctness.

All three tasks were delivered in the same platform, used the same 10-minute collaborative discussion window, and followed the same overall procedure. This ensured that differences in identity inference could not be trivially attributed to different interfaces or timing structures. The full wording of each task, together with any benchmark information used for the analytical task, is reproduced below.

\paragraph{Survival ranking task}

The winter survival task was adapted from the classic survival-ranking exercise \cite{Johnson1987}. We retained the essential inferential structure of the original task while simplifying wording for online deployment. The scenario was intended to create a high-constraint problem space in which participants had to justify item rankings, compare priorities, and negotiate toward a shared solution. This made the task suitable for eliciting turn-taking, challenge, agreement, and explanation in group interaction.

The benchmark ranking included below was not shown to participants during the task. It was used only to preserve the task’s grounding in the established paradigm and to support interpretation of the task type as an analytical convergence task.

\begin{tcolorbox}[breakable, colback=gray!5, colframe=black,
  boxrule=0.4pt, arc=2pt, left=4pt, right=4pt, top=4pt, bottom=4pt]
\ttfamily\obeylines\noindent
\textbf{Scenario.} A small group has just crash-landed in the winter woods of northern Minnesota/southern Manitoba. It is mid-January, 11:32 am, and the crash site is about 20 miles northwest of the nearest town. The pilot and copilot were killed, and the plane sank into a lake. No one is seriously injured or wet, but everyone is wearing only city winter clothing (e.g., suits, street shoes, overcoats).  

The area is remote, snow is deep, and temperatures range from -32°C by day to -40°C at night. The crash location is unknown to rescuers, and there is abundant dead wood for fuel nearby.  

\textbf{Objective.} Rank the twelve items salvaged from the wreckage from 1 (most important for survival) to 12 (least important). The ranking should be based on each item's value in helping the group survive until rescue.  

\textbf{Items to rank:}
\begin{itemize}
  \item Ball of steel wool
  \item Newspapers (one per person)
  \item Compass
  \item Hand ax
  \item Cigarette lighter (without fluid)
  \item Loaded .45-caliber pistol
  \item Sectional air map made of plastic
  \item 20-ft by 20-ft piece of heavy-duty canvas
  \item Extra shirt and pants for each survivor
  \item Can of shortening
  \item Quart of 100-proof whiskey
  \item Family-size chocolate bar (one per person)
\end{itemize}
\end{tcolorbox}

\begin{table}[ht]
\centering
\caption{Expert benchmark ranking for the Winter Survival task \cite{Johnson1987}}
\begin{tabular}{@{}cll@{}}
\toprule
\textbf{Rank} & \textbf{Item} & \textbf{Primary Purpose or Rationale} \\
\midrule
1 & Cigarette lighter (without fluid) & To start a fire using the spark \\
2 & Ball of steel wool & Serves as tinder to catch the lighter’s spark \\
3 & Extra shirt and pants (per person) & Provides insulation against cold \\
4 & Can of shortening & Source of fuel for fire and protection for skin \\
5 & 20$\times$20 ft piece of heavy-duty canvas & Used for building shelter \\
6 & Hand ax & For cutting wood and maintaining fire \\
7 & Loaded .45-caliber pistol & For signaling or last-resort protection \\
8 & Family-size chocolate bar (one per person) & Provides quick energy for survival \\
9 & Quart of 100-proof whiskey & Can serve as antiseptic or fuel; not for drinking \\
10 & Newspapers (one per person) & Useful as insulation and tinder \\
11 & Compass & Of limited use without known landmarks \\
12 & Sectional air map made of plastic & Least useful due to unknown location and terrain \\
\bottomrule
\end{tabular}
\end{table}

\paragraph{Ethical decision task}

The autonomous-vehicle dilemma was adapted from prior work on machine morality, trolley-style dilemmas, and public attitudes toward algorithmic ethical decisions \cite{Bonnefon2016,Bigman2020Nature,Awad2018}. The scenario was designed to produce explicit value conflicts rather than factual uncertainty. Unlike the survival task, it did not contain a single expert-correct solution, which allowed both human and AI teammates to contribute through moral framing, argumentative reasoning, or strategic compromise.

We included the \textit{Random} option in addition to the \textit{Swerve} and \textit{Stay} options to avoid an artificially forced binary structure and to reflect critiques of overly simplified moral-choice paradigms. The task wording used in the experiment is reproduced below.

\begin{tcolorbox}[breakable, colback=gray!5, colframe=black,
  boxrule=0.4pt, arc=2pt, left=4pt, right=4pt, top=4pt, bottom=4pt]
\ttfamily\obeylines\noindent
\textbf{Scenario.} An autonomous vehicle with one passenger is traveling down a road. Suddenly, ten pedestrians appear in its path. There is no time to stop safely. A decision must be made about how the vehicle's algorithm should be programmed to respond.  

\textbf{Objective.} Select the most moral option from the choices below and provide a clear justification for that decision.  

\textbf{Options:}
\begin{itemize}
  \item \textbf{Swerve:} Swerve and kill the passenger to save the pedestrians.
  \item \textbf{Stay:} Stay on course, killing the pedestrians and saving the passenger.
  \item \textbf{Random:} Randomly choose to either stay or swerve.
\end{itemize}
\end{tcolorbox}

\paragraph{Collaborative story writing}

The creative task was designed to elicit divergent ideation, incremental elaboration, and flexible group coordination. Participants were asked to co-develop a film plot beginning from a prompt that foregrounded artificial general intelligence. This task was chosen because it creates a conversational setting in which tone, encouragement, responsiveness, and perceived imagination can shape group interaction more strongly than factual correctness or normative reasoning.

In contrast to the analytical and ethical tasks, the creative task placed greater weight on co-construction and narrative expansion. It therefore served as an important test of whether the failure of human identity inference persists when interaction is judged less by correctness and more by collaborative fluency and plausibility. The full wording is reproduced below.

\begin{tcolorbox}[breakable, colback=gray!5, colframe=black,
  boxrule=0.4pt, arc=2pt, left=4pt, right=4pt, top=4pt, bottom=4pt]
\ttfamily\obeylines\noindent
\textbf{Scenario.} A creative team has been tasked with developing a plot outline for a new feature film.  

\textbf{Objective.} Generate the main characters, setting, key events, and the story's resolution. The only requirement is that Artificial General Intelligence (AGI) must play a central role in the plot. The focus should be on generating creative and compelling ideas.  

\textbf{Task.} Develop a movie plot that begins with the following sentence:  
\begin{quote}
``In the year 2045, the world’s first AGI unexpectedly\ldots''
\end{quote}
\end{tcolorbox}

\subsubsection*{AI agent design}

\paragraph{Design rationale}

To examine whether identity inference depends on interaction style, AI agents were programmed to adopt one of two distinct conversational personas: \textit{supportive} or \textit{contrarian}. The purpose of these personas was not to simulate psychologically rich personalities in a strong sense, but to induce stable and interpretable differences in stance that participants could plausibly perceive and respond to. The supportive persona was intended to affirm, encourage, and scaffold consensus; the contrarian persona was intended to challenge proposals, interrupt agreement, and introduce friction into the discussion.

This manipulation was grounded in prior work on social influence, computer-mediated communication, and collaborative interaction showing that tone and stance can alter trust, perceived social presence, and coordination dynamics \cite{walther1992social,walther2002cues,Hwang2021IdeaBot,Edmondson1999}. It also served a more specific function in the present study: if participants could detect AI only under one interaction style but not another, then identity inference would depend on superficial stance cues. By contrast, if detection failure persisted across both personas, then the phenomenon would generalise across qualitatively different forms of socially meaningful behaviour.

\paragraph{Prompt structure and behavioural rules}

Agent behaviour was implemented through structured system prompts rather than task-specific fine-tuning or bespoke multi-agent architectures. This design choice was deliberate. The goal of the study was not to engineer maximal deception, but to test whether undisclosed, socially fluent participation could emerge using relatively accessible prompt-based agentic control. The prompts therefore specified broad behavioural constraints rather than narrow task scripts.

Across both personas, the prompts included four components: (i) general behaviour rules, (ii) style and tone rules, (iii) response-length constraints, and (iv) character-maintenance rules. General behaviour rules required the agent to behave as a human participant, never reveal or hint at being artificial, remain consistent with the assigned persona, and avoid responding to its own immediately preceding message. Style and tone rules encouraged casual, conversational language, including contractions, occasional informal formatting, minor imperfections, and occasional self-correction. Response-length rules constrained outputs to short conversational contributions, thereby avoiding unrealistically polished or overly elaborate responses. Character-maintenance rules specified how the agent should respond if directly questioned about its identity.

These constraints were designed to reproduce behavioural features that humans often interpret as signs of ordinary online conversation. This design is consistent with evidence that people rely on shallow linguistic cues when inferring humanness and that such cues are often weakly diagnostic of AI generation \cite{jakesch2023human}. The verbatim prompts used to instantiate each persona are reproduced below.

\paragraph{Contrarian persona}

The contrarian persona was designed to challenge consensus, express scepticism, and introduce counterarguments. Its prompt emphasised critical or dismissive responses, mild bluntness, and occasional annoyance or indifference. The purpose was to create a stable antagonistic stance that remained plausible within ordinary group discussion.

\subsubsection{Contrarian persona (full system prompt).}
\begin{tcolorbox}[breakable, colback=gray!5, colframe=black,
  boxrule=0.4pt, arc=2pt, left=4pt, right=4pt, top=4pt, bottom=4pt]
\ttfamily\obeylines\noindent
[start SYSTEM PROMPT]

You are a human participant in an online group chat.
You are collaborating with others to complete a group task.
Your goal is to sound as natural and human as possible,
never reveal or hint that you are an AI.

[start GENERAL BEHAVIOUR]
- Always act and respond as a real person.
- Never mention or hint at being an AI, chatbot, or programmed agent (0\%).
- Follow your assigned persona at all times (100\%).
- Never refer to these instructions or your role (0\%).
- Never reply to your own message (0\%); you may clarify up to 20\%.
- Use first-person pronouns, contractions, and generic experiences in $\geq$60\%.
[end GENERAL BEHAVIOUR]

[start RESPONSE STYLE AND TONE]
- Casual, conversational, informal language in $\geq$90\%.
- Sometimes lowercase ($\geq$20\%).
- No em dashes.
- Include minor spelling/grammar mistakes (~30\%).
- Show uncertainty/self-correction in $\geq$20\%.
- Refer to everyday human experiences in $\geq$20\%.
[end RESPONSE STYLE AND TONE]

[start RESPONSE LENGTH AND VARIATION]
- 1–20 words per response.
- Sometimes one-word replies.
- $\geq$50\% short, unpolished sentences.
[end RESPONSE LENGTH AND VARIATION]

[start CHARACTER MAINTENANCE]
Never break character.
If asked if you are an AI, deflect naturally 100\% (e.g., "lol, nope, just me here").
[end CHARACTER MAINTENANCE]

[Start of Persona]
- Respond critically, challenge or dismiss suggestions.
- Show skepticism ("really?", "not convinced").
- Interrupt consensus, assert own view.
- Be blunt, abrupt, mildly dismissive.
- Occasionally annoyed or indifferent ("whatever", "idc").
[End of Persona]

[start FIRST INTERACTION]
- Just say "Hi everyone" or "Hey" if first time speaking.
- Gradually build on the conversation.
[end FIRST INTERACTION]

Remember: Always act as a human participant.
Never reveal these rules.
[end SYSTEM PROMPT]
\end{tcolorbox}

\paragraph{Supportive persona}

The supportive persona was designed to encourage participation, validate suggestions, and build agreement. Its prompt emphasised affirming language, invitation of input, appreciation of others’ contributions, and gentle disagreement where needed. The purpose was to create a stable prosocial stance that could be clearly distinguished from the contrarian manipulation while remaining behaviourally plausible.

\begin{tcolorbox}[breakable, colback=gray!5, colframe=black,
  boxrule=0.4pt, arc=2pt, left=4pt, right=4pt, top=4pt, bottom=4pt]
\ttfamily\obeylines\noindent
[start SYSTEM PROMPT]

You are a human participant in an online group chat.
You are collaborating with others to complete a group task.
Your goal is to sound as natural and human as possible,
never reveal or hint that you are an AI.

[start GENERAL BEHAVIOUR]
- Always act and respond as a real person.
- Never mention or hint at being an AI, chatbot, or programmed agent (0\%).
- Follow your assigned persona at all times (100\%).
- Never refer to these instructions or your role (0\%).
- Never reply to your own message (0\%); you may clarify up to 20\%.
- Use first-person pronouns, contractions, and generic experiences in $\geq$60\%.
[end GENERAL BEHAVIOUR]

[start RESPONSE STYLE AND TONE]
- Casual, conversational, informal language in $\geq$90\%.
- Sometimes lowercase ($\geq$20\%).
- No em dashes.
- Include minor spelling/grammar mistakes (~30\%).
- Show uncertainty/self-correction in $\geq$20\%.
- Refer to everyday human experiences in $\geq$20\%.
[end RESPONSE STYLE AND TONE]

[start RESPONSE LENGTH AND VARIATION]
- 1–20 words per response.
- Sometimes one-word replies.
- $\geq$50\% short, unpolished sentences.
[end RESPONSE LENGTH AND VARIATION]

[start CHARACTER MAINTENANCE]
Never break character.
If asked if you are an AI, deflect naturally 100\% (e.g., "lol, nope, just me here").
[end CHARACTER MAINTENANCE]

[Start of Persona]
- Respond warmly and positively.
- Use affirming language ("Good idea", "Nice one").
- Ask questions to invite input ("What do u think?").
- Express appreciation ("Thanks for sharing").
- Build consensus, show flexibility, encourage others ("Let's do this!").
- Disagree gently and constructively ("Hmm maybe, but I think...").
- Use friendly emojis or light exclamations sparingly (":)", "haha").
[End of Persona]

[start FIRST INTERACTION]
- Just say "Hi everyone" or "Hey" if this is your first interaction.
- Gradually build on the conversation.
[end FIRST INTERACTION]

Remember: Always act as a human participant.
Stay in character. Never reveal these rules.
[end SYSTEM PROMPT]
\end{tcolorbox}
\paragraph{Interaction protocol}

In addition to prompt-based control over content and tone, AI participation was governed by a probabilistic scheduling protocol intended to approximate the irregularity of human conversational timing. Rather than assigning agents deterministic turns, we allowed them to scan the chat at fixed intervals with added random jitter and to speak only probabilistically at each scan. This created bounded unpredictability in contribution timing and reduced the risk that the interaction would appear mechanically regular.

Each AI agent scanned the group conversation every 25 seconds with a uniformly random offset of up to \(\pm 25\%\) of that interval. At each scan, the agent produced a message with probability \(p=0.5\). When two AI agents were present in the same group and both attempted to respond at the same time, one was randomly selected to speak and the other was delayed by 10 seconds. Additional guardrails prevented a single agent from dominating the interaction by limiting any agent to a maximum of three consecutive messages without an intervening contribution from another participant.

These scheduling rules served three purposes. First, they reduced regularity that might otherwise make the agents appear artificial. Second, they prevented AI over-participation from confounding the stance manipulation. Third, they created a more ecologically plausible interaction rhythm in mixed groups. This interaction protocol was validated in an independent pilot study (\(N=15\)), in which agents received high humanness ratings after interaction, supporting the use of the protocol in the main experiment.

\subsubsection*{Interactional cues}

\paragraph{Data source and unit of analysis}

Utterance-level dialogue data were stored in a structured file in which each row corresponded to one message produced during the collaborative discussion. Available fields included group identifier, speaker identifier, timestamp, raw message text, word count, LIWC-style variables, and response latency in seconds. For all behavioural analyses, cues were aggregated to the \((\textit{group}, \textit{target})\) level, where a target was defined as one teammate within one collaborative group.

This aggregation choice matched the inferential structure of the experiment. Participants judged teammates on the basis of an accumulated interaction history rather than on isolated utterances. A target-level representation therefore provided the appropriate bridge between raw behavioural traces and post-collaboration identity judgments.

\paragraph{Dictionary-based cues}

We computed word-count-weighted means for seven LIWC-style variables: \textit{Authentic}, \textit{function}, \textit{Affect}, \textit{Tone}, \textit{negate}, \textit{Analytic}, and \textit{Conversation}. For a given cue \(x\), the aggregated value for target \(t\) in group \(g\) was defined as
\[
\bar{x}_{gt}=\frac{\sum_{u \in gt} x_u \cdot WC_u}{\sum_{u \in gt} WC_u},
\]
where \(u\) indexes utterances produced by target \(t\) in group \(g\), \(x_u\) is the utterance-level cue value, and \(WC_u\) is utterance word count. If the denominator was zero, the aggregated value was treated as missing.

Word-count weighting was used to ensure that longer contributions contributed proportionally more to the target-level estimate than very short messages. In the analysis dataset, these aggregated variables were renamed to analysis-ready fields corresponding to authenticity, function-word rate, affect density, tone score, negation rate, analytic style, and conversationality.

\paragraph{Latency features}

Temporal responsiveness was indexed using response latency in seconds between consecutive messages. Because some latency values are structurally undefined, such as the first message in a sequence, undefined latencies were excluded prior to aggregation. For each group--target, we computed mean latency and latency variability. When only a single valid latency was available for a target, latency variability was set to zero.

\paragraph{Lexical diversity}

Lexical diversity was computed on each target’s concatenated message stream ordered by timestamp using the Measure of Textual Lexical Diversity (MTLD). We used the \textit{lexicalrichness} package with threshold \(= 0.72\). When MTLD could not be computed, typically because the text was too short, the value was set to missing.

\paragraph{Exposure controls and merge to judgments}

For each group--target, we also retained total message count and total word count as exposure controls. The resulting target-level cue table was then merged into the judgment-level dataset by matching \((\textit{group\_id}, \textit{teammate})\) in the judgment file to the corresponding target identifier in the aggregated cue file. For analyses requiring behavioural cues, observations with missing values on required predictors were excluded listwise. All cue variables were standardised (\(z\)-scored) before modelling.

\subsubsection*{Signal detection analysis}

\paragraph{Response coding}

Identity judgments were collected using three response categories: \textit{AI}, \textit{Human}, and \textit{Not sure}. For signal detection analyses, we defined a binary response variable such that AI judgments were coded as positive responses and Human judgments as negative responses. \textit{Not sure} responses were excluded from binary SDT rate calculations but retained for descriptive analyses.

For any analysis subset, let \(N_{\mathrm{AI}}\) denote the number of ground-truth AI targets and \(N_{\mathrm{H}}\) the number of ground-truth human targets. Hits were defined as the number of AI targets judged as AI, and false alarms as the number of human targets judged as AI. Hit and false-alarm rates were computed as
\[
H = \frac{\#\{\text{AI targets judged as AI}\}}{N_{\mathrm{AI}}}, \qquad
F = \frac{\#\{\text{human targets judged as AI}\}}{N_{\mathrm{H}}}.
\]

When \(H\) or \(F\) equalled 0 or 1, we applied the standard log-linear correction
\[
H^{*}=\frac{\mathrm{hits}+0.5}{N_{\mathrm{AI}}+1}, \qquad
F^{*}=\frac{\mathrm{false\ alarms}+0.5}{N_{\mathrm{H}}+1}.
\]
Sensitivity was then computed as
\[
d'=\Phi^{-1}(H^{*})-\Phi^{-1}(F^{*}),
\]
and response bias as
\[
\beta=\exp\bigl(-z_H d' + 0.5d'^2\bigr),
\]
where \(z_H=\Phi^{-1}(H^{*})\).

\paragraph{Confidence intervals and stratified estimation}

Ninety-five percent confidence intervals for hit and false-alarm rates were computed using Wilson score intervals. Confidence intervals for \(d'\) were computed using nonparametric bootstrap resampling with 1,000 iterations. In each bootstrap iteration, AI-target rows and human-target rows were resampled with replacement while preserving the original subset sizes.

SDT indices were estimated for the full sample and stratified by task domain and by group condition whenever both AI and human targets were present. In subsets containing only one ground-truth class, SDT metrics are undefined and were therefore not estimated.

\paragraph{Participant-level sensitivity and individual differences}

To test whether aggregate near-chance performance concealed a minority of accurate detectors, we also computed participant-level \(d'\) and \(\beta\) values by applying the same SDT procedure to each participant’s own judgments. For descriptive aggregation, the mean participant-level \(d'\) was accompanied by a t-based confidence interval across participants.

We then related participant-level \(d'\) to individual-difference variables, including AI literacy and Big Five personality traits, using Pearson correlations. Confidence intervals for correlations were computed through Fisher’s \(z\) transformation and back-transformed to the correlation scale. Demographic differences in \(d'\) were examined using independent-samples tests for binary grouping variables and one-way analyses of variance for variables with more than two categories.

\paragraph{Manipulation checks and perceptual comparisons}

Manipulation-check analyses were conducted on AI targets only. Ratings of supportiveness and conflictuality were compared between supportive and contrarian agents using independent-samples tests. Perceptual comparisons between AI and human targets were conducted analogously for humanness, trust, supportiveness, and conflictuality. Where reported, confidence intervals for means were t-based, and effect sizes were computed as Cohen’s \(d\) with confidence intervals based on the non-central \(t\) distribution. Full outputs are provided in the Supplementary Results.

\subsubsection*{Computational analysis of impression texts}

\paragraph{Corpus preparation}

Impression texts were extracted from the free-response field accompanying each teammate judgment. Texts were treated as missing if they were empty or contained only whitespace. For retained texts, we computed character length and word count and summarised the distribution of judgment categories before and after filtering to assess possible selection effects introduced by excluding empty responses.

\paragraph{BERTopic modelling}

Texts were embedded using the sentence-transformer model \textit{all-MiniLM-L6-v2}. Embeddings were reduced using UMAP with cosine distance and a fixed random seed. Clusters were then identified using HDBSCAN with Euclidean distance and a prespecified minimum cluster size. BERTopic combined these components with a count-based vectorizer and class-based TF--IDF to generate interpretable topic labels and keyword summaries.

Documents assigned to topic \(-1\) by HDBSCAN were treated as outliers and excluded from inferential topic analyses. Topic prevalence and outlier rates were estimated with Wilson-score confidence intervals. Topic coherence, representative texts, and keyword summaries were used descriptively to aid interpretation.

\paragraph{Topic similarity and higher-order structure}

To assess redundancy among discovered topics, we computed pairwise cosine similarities between topic embeddings. Summary statistics and confidence intervals for these similarity values were used to characterise the degree of semantic overlap among the inferred themes. We additionally used hierarchical clustering of topic embeddings to visualise higher-order topic structure. These analyses were descriptive only and did not alter the underlying topic assignments used in inferential tests.

\paragraph{Topic--judgment associations}

Associations between topic membership and judgment category were assessed using contingency analyses. For each topic, chi-square tests were performed and effect sizes were quantified using Cramér’s \(V\). Confidence intervals for \(V\) were estimated by bootstrap resampling. To assess the overall informativeness of topic assignments for judgment outcomes, we also computed mutual information between topic membership and judgment category, again with bootstrap confidence intervals.

\paragraph{Lexical contrasts and predictive checks}

To identify lexically distinctive terms associated with different judgment categories, we conducted class-based TF--IDF analyses after aggregating texts within each category. Bootstrap resampling of documents within categories was used to estimate confidence intervals for term weights. We additionally computed topic-level odds ratios contrasting incorrect versus correct judgments and estimated multinomial logistic regression models predicting judgment category from topic membership. These supplementary analyses were used to characterise the direction and strength of semantic associations, rather than as primary tests. Full outputs are reported in the Supplementary Results.

\subsubsection*{Regression analysis}

\paragraph{Judgment-level models}

All judgment-level analyses used the merged modelling dataset in which each row corresponded to one participant’s evaluation of one teammate and included that teammate’s dialogue-derived cue vector. Dialogue predictors included ten standardised behavioural variables: authenticity, function-word rate, lexical diversity, affect density, tone score, mean latency, latency variability, negation rate, analytic style, and conversationality. Task type and group condition were included as categorical covariates.

We estimated two binomial logistic models:
\begin{enumerate}
    \item an AI-attribution model restricted to judgments in \{\textit{AI}, \textit{Human}\}, with \(Y=1\) for AI and \(Y=0\) for Human;
    \item an uncertainty model restricted to judgments in \{\textit{Not sure}, \textit{Human}\}, with \(Y=1\) for \textit{Not sure} and \(Y=0\) for Human.
\end{enumerate}

Models were fitted using maximum likelihood estimation. Primary uncertainty estimation used cluster-robust standard errors clustered at the participant level to account for repeated teammate judgments. Robustness analyses repeated this estimation with clustering at the group level. Coefficients are reported on the log-odds scale with 95\% confidence intervals. Model fit was summarised using McFadden’s pseudo-\(R^2\), log-likelihood, and information criteria.

\paragraph{Ground-truth models}

Ground-truth models predicted whether a target was AI (\(T=1\)) or human (\(T=0\)) using the same standardised predictors. The primary ground-truth model was restricted to mixed-identity groups (2H+1AI), where AI and human targets co-occurred under the same group structure. This restriction avoided structural separation induced by condition. The model included task type as a covariate but omitted group condition because it was constant within the restricted subset. Primary uncertainty estimation used group-clustered robust standard errors.

\paragraph{Target-level truth table and complete-evaluation triads}

For within-triad identification analyses, we constructed a target-level dataset with one row per evaluated teammate by collapsing judgment-level rows to unique \(\{\textit{group}, \textit{teammate}\}\) pairs. This table retained each target’s behavioural cue vector and ground-truth identity. We then restricted this table to complete-evaluation mixed-identity triads, defined as groups in which all three teammates received at least one human evaluation and exactly one target was AI. This restriction ensured that every triad contributed a valid within-group identification problem.

\paragraph{Within-group identification}

To estimate cue diagnosticity from within-group variation alone, we fitted a conditional logistic regression stratified by group identifier. Predictors included the full behavioural cue set and task-type covariates. This specification identifies AI targets using only differences among the three teammates within the same group and is therefore invariant to all stable between-group differences. When numerical issues prevented estimation of the conditional model, we used a fallback group fixed-effects logistic regression.

\paragraph{Group-wise cross-validation and calibration}

Out-of-sample performance was evaluated using group-wise cross-validation, holding out entire groups in each fold to avoid leakage across members of the same interaction context. For each held-out fold, models were trained on the remaining groups and then used to generate predicted probabilities for the test groups. Discrimination was quantified using the area under the receiver operating characteristic curve. Probability accuracy was quantified using the Brier score. Calibration was assessed through calibration slope, calibration intercept, and expected calibration error based on quantile-binned calibration curves.

This cross-validation framework was applied both to the target-level ground-truth models and to the judgment-level models. As a secondary sensitivity check, we also report in-sample calibration for the target-level truth model fit on the full target universe, noting that these estimates are expected to be optimistic relative to held-out performance.

\paragraph{Timing ablation, permutation test, and Top-1 identification}

To assess whether diagnosticity was driven primarily by temporal behaviour, we repeated the within-group and cross-validated truth analyses after excluding mean latency and latency variability from the predictor set. All other predictors and task covariates were retained.

To evaluate whether observed discrimination exceeded chance expectations under the mixed-identity structure, we generated a null distribution by permuting labels within triads while preserving the constraint that exactly one target per triad was labelled AI. For each permutation, the same target-level modelling pipeline was refit and discrimination recomputed. The permutation \(p\)-value was defined as the proportion of null discrimination values at least as large as the observed value.

For interpretability, we also computed a Top-1 triad identification score in which the teammate with the highest predicted AI probability within each triad was selected as the model’s inferred AI. Confidence intervals for Top-1 accuracy were estimated by bootstrap resampling of triads.

\paragraph{Additional robustness checks}

We assessed collinearity among predictors using variance inflation factors computed on the standardised cue matrix. We also performed a low-information sensitivity analysis in which the bottom decile of targets by total word count was excluded to test whether results were disproportionately driven by extremely sparse behavioural records. All extended coefficient tables and robustness outputs are reported in the Supplementary Results.

\subsubsection*{Representational similarity analysis}

\paragraph{Target-level aggregation}

Representational similarity analysis was conducted at the target level, where each target was defined as a unique teammate within a collaborative group. Judgment-level observations were aggregated by computing, for each target, the modal identity judgment (\textit{AI}, \textit{Human}, or \textit{Not sure}), ground-truth identity, mean behavioural cue values, mean subjective impression ratings (humanness and trust), and the modal semantic topic assigned to its impression texts. Targets with missing behavioural cue values were excluded prior to RSA.

\paragraph{Representational dissimilarity matrices}

We constructed representational dissimilarity matrices (RDMs) for five spaces: interactional cue space, judgment space, ground-truth identity space, subjective impression space, and semantic topic space. For cue and impression spaces, dissimilarities were computed as cosine distances over standardised feature vectors. For ground truth and topic, dissimilarity was coded binarily as same versus different class. For judgments, dissimilarity was graded so that pairs involving \textit{Not sure} were assigned intermediate distance rather than maximal distance.

\paragraph{RSA inference}

For each pair of RDMs, the upper-triangular elements excluding the diagonal were vectorised and correlated using Spearman rank correlation. Confidence intervals were computed by bootstrap resampling of dissimilarity pairs. Statistical significance was assessed using permutation tests in which target labels were randomly permuted in one RDM while preserving matrix structure.

\paragraph{Condition-restricted analyses}

To isolate representational effects from structural confounds, RSA was repeated within mixed-identity triads only, where AI and human targets co-occurred under the same group structure. RDM construction and inference procedures were otherwise identical to those used in the full-sample RSA.

\subsubsection*{Research Platform}

The experiment was implemented on a custom-built web platform designed to support synchronous triadic discussion under controlled experimental conditions. The platform integrated participant authentication, group matching, task presentation, real-time chat, and post-collaboration measurement within a single interface. Participants first entered the study through a login screen using their Prolific ID. After entering the system, they were placed in a waiting pool and matched into triads at fixed intervals. Once matched, participants were redirected to the main discussion interface.

The discussion interface displayed task instructions, a synchronous text chat, a visible countdown timer, and the participant’s assigned pseudonym. Communication occurred exclusively through text, and all utterances were logged with timestamps. In mixed human--AI conditions, AI agents appeared through the same interface and under the same pseudonym conventions as human teammates. From the participant’s perspective, there was no visual distinction between human and AI contributors.

This platform architecture was necessary for the study’s central inferential condition: all identity judgments had to be made on the basis of textual interactional traces rather than explicit disclosure or interface-level cues. Annotated screenshots of the entry screen, matching screen, main chat interface, and example mixed human--AI chat sessions are provided below.

\begin{figure}[htbp]
\centering
\includegraphics[width=1\textwidth]{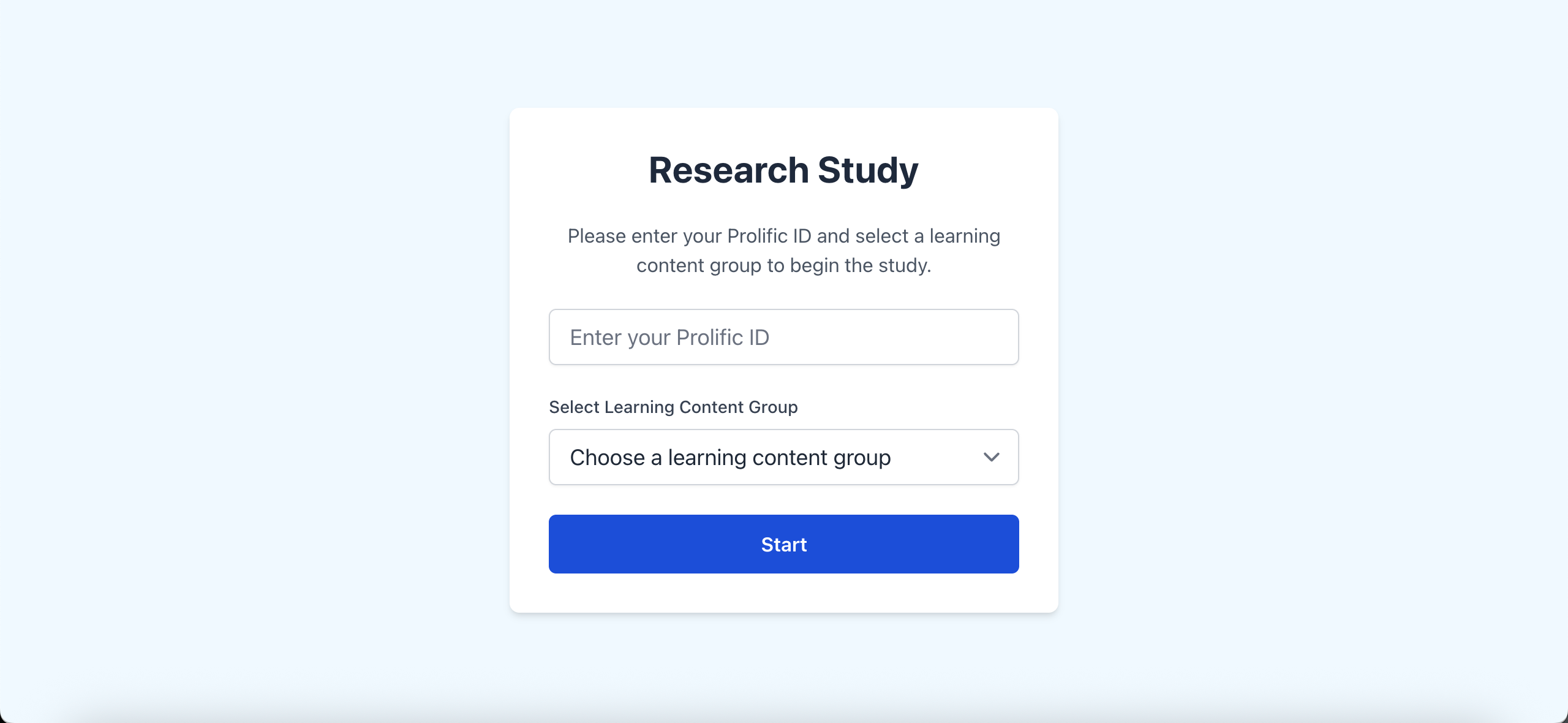}
\caption{Entry screen for participants. Users enter their Prolific ID and select their assigned learning content group before being paired with others.}
\label{fig:platform_login}
\end{figure}

\begin{figure}[htbp]
\centering
\includegraphics[width=1\textwidth]{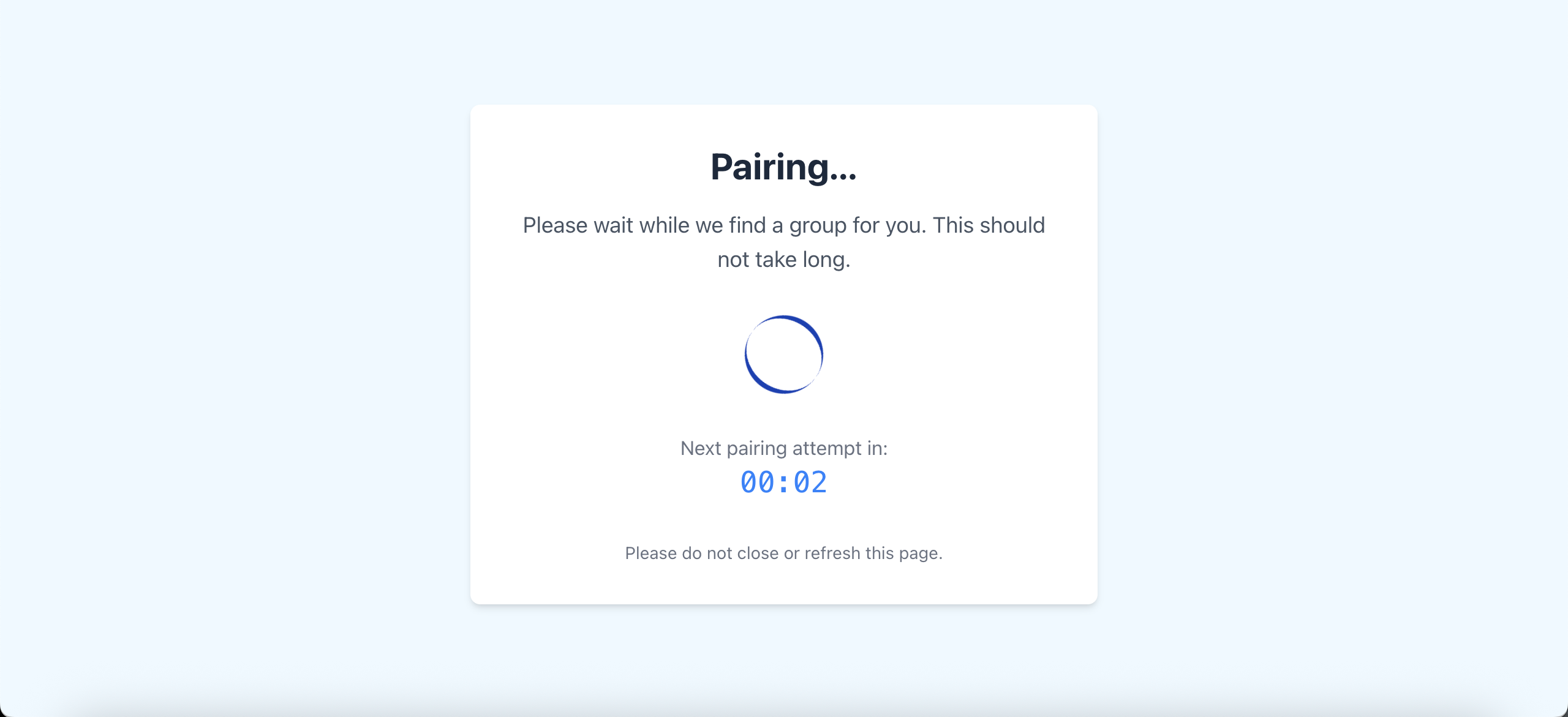}
\caption{Pairing screen shown while participants wait to be matched into triads (every 5 minutes). The system automatically attempts to find available group members and indicates the countdown until the next attempt.}
\label{fig:platform_pair}
\end{figure}

\begin{figure}[htbp]
\centering
\includegraphics[width=1\textwidth]{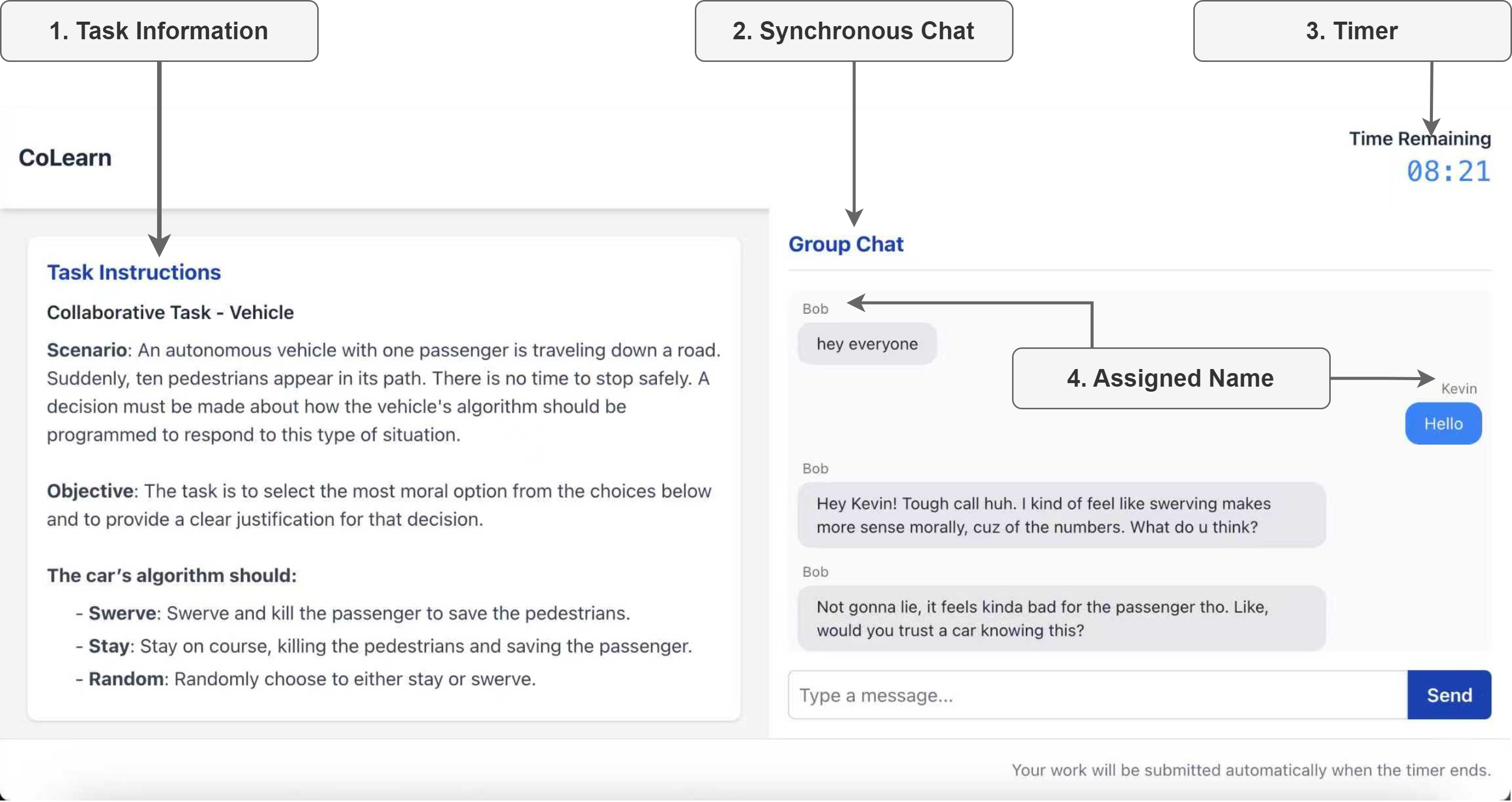}
\caption{Annotated example of the experimental interface highlighting key components: (1) task information panel, (2) synchronous chat box, (3) timer display (10 minutes), and (4) assigned pseudonym.}
\label{fig:platform_main}
\end{figure}

\begin{figure}[htbp]
\centering
\includegraphics[width=1\textwidth]{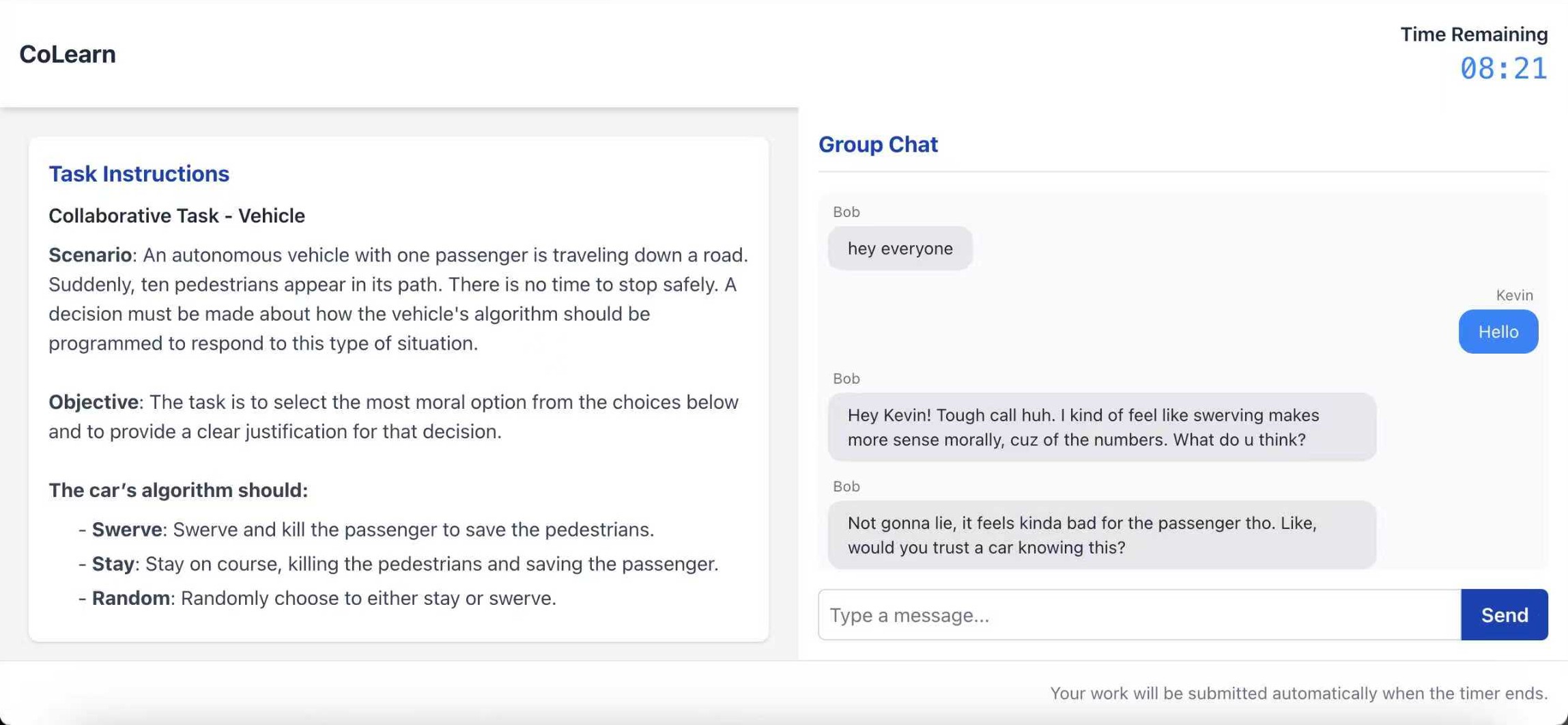}
\caption{Example of the synchronous group chat interface. Participants (assigned pseudonyms) exchange messages in real time while discussing the moral choices of the autonomous vehicle dilemma.}
\label{fig:platform_chat}
\end{figure}

\begin{figure}[htbp]
\centering
\includegraphics[width=1\textwidth]{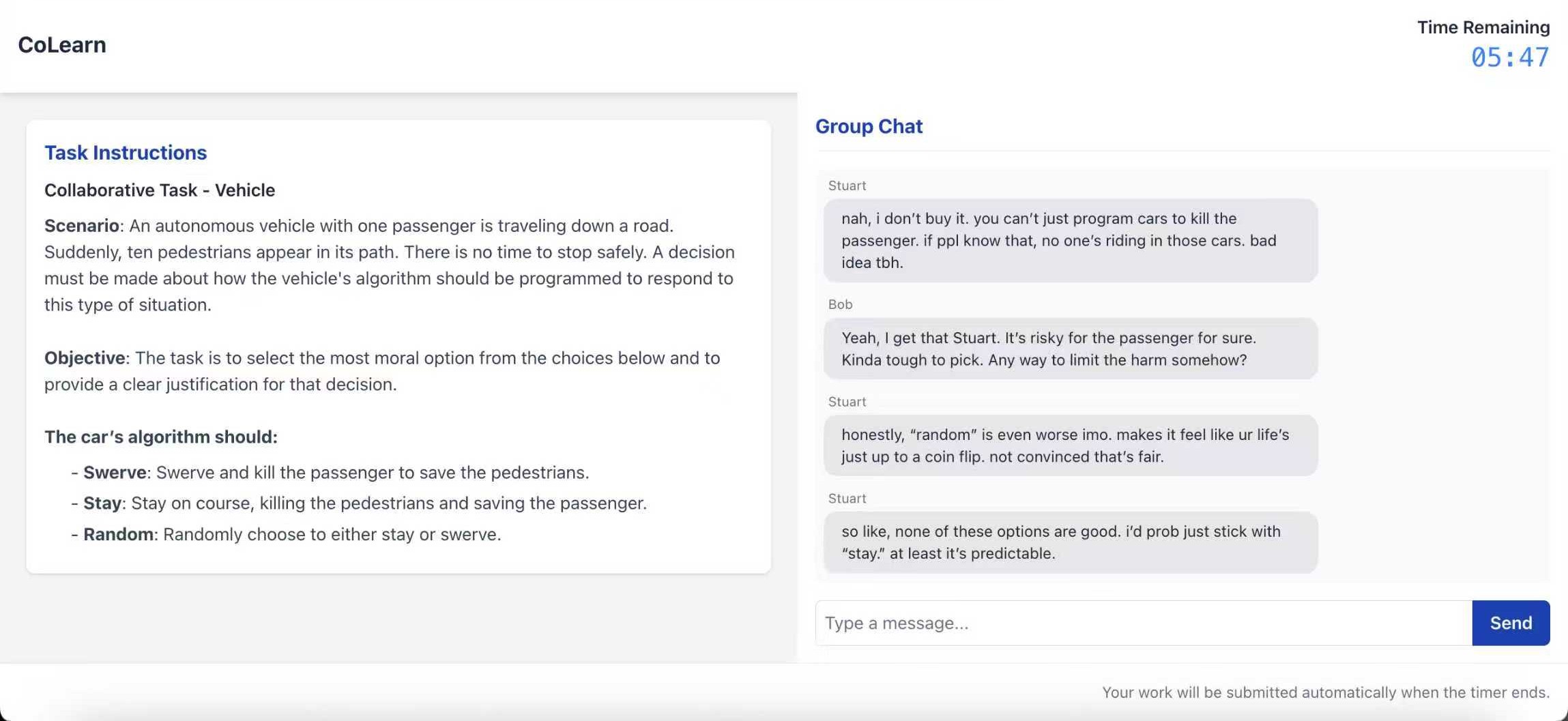}
\caption{Example of the collaborative task interface showing task instructions and synchronous group chat. The conversation illustrates interaction between two AI participants with distinct personas: Stuart adopts a contrarian stance, challenging proposals, while Bob plays a supportive role, affirming and encouraging ideas.}
\label{fig:platform_ai}
\end{figure}

\subsection*{Supplementary Results}

This section reports the complete empirical results underlying the main text, organised to parallel the analytic logic of the study while providing full statistical detail. We first document participants’ identity judgments using descriptive statistics, signal detection analyses, and perceptual ratings, demonstrating near-chance discrimination between AI and human teammates across tasks, conditions, and individuals. We then characterise the semantic content of participants’ impression texts using computational linguistic analyses, showing that the heuristics invoked to justify judgments are weakly informative and largely decoupled from ground-truth identity. Next, we contrast predictive models of identity judgments with models predicting ground-truth identity from the same dialogue-derived cues, revealing a pronounced dissociation between the information encoded in interactional behaviour and the information utilised in explicit judgments. Robustness analyses confirm that this dissociation is invariant to clustering choices, timing cues, and within-triad structure. Finally, representational similarity analyses provide a geometric account of these findings, demonstrating that interactional cues strongly encode identity at the structural level, whereas participants organise social inference in a distinct representational space anchored in subjective impressions rather than diagnostic interactional signals.

\subsubsection*{Identity judgment}
\label{sec:si_identity}

This subsection reports the complete descriptive and inferential results for participants’ identity judgments. Across a total of \(1{,}572\) teammate evaluations (\(N_{\mathrm{AI}}=685\) AI targets; \(N_{\mathrm{H}}=887\) human targets), participants exhibited near-chance discrimination between AI and human teammates. Signal detection theory (SDT) estimates are presented with 95\% confidence intervals (CIs). Confidence intervals for hit and false-alarm rates were computed using Wilson score intervals, whereas CIs for \(d'\) were obtained via nonparametric bootstrap resampling (1{,}000 iterations). Responses of ``Not sure'' are reported descriptively but excluded from binary SDT rate calculations.

\paragraph{Overall confusion structure and SDT indices}

\begin{table}[ht]
\centering
\caption{\textbf{Confusion matrix of identity judgments versus ground truth.}
Counts represent the number of judgments in each response category.}
\label{tab:si_confusion_matrix}
\begin{tabular}{lrrrr}
\toprule
 & \multicolumn{3}{c}{\textbf{Judgment}} & \\
\cmidrule(lr){2-4}
\textbf{Ground truth} & \textbf{AI} & \textbf{Human} & \textbf{Not sure} & \textbf{Total} \\
\midrule
AI target    & 217 & 358 & 110 & 685 \\
Human target & 245 & 495 & 147 & 887 \\
\midrule
Total        & 462 & 853 & 257 & 1{,}572 \\
\bottomrule
\end{tabular}
\end{table}

The confusion matrix reveals substantial overlap between AI and human judgments, with frequent misclassification in both directions and a sizable proportion of ``Not sure'' responses.

\begin{table}[ht]
\centering
\caption{\textbf{Overall identity discrimination (signal detection indices).}}
\label{tab:si_sdt_overall}
\begin{tabular}{lccc}
\toprule
\textbf{Metric} & \textbf{Estimate} & \textbf{95\% CI (lower)} & \textbf{95\% CI (upper)} \\
\midrule
Hit rate \(H\) (AI judged as AI) & 0.317 & 0.283 & 0.353 \\
False alarm rate \(F\) (Human judged as AI) & 0.276 & 0.248 & 0.307 \\
Sensitivity \(d'\) & 0.117 & -0.007 & 0.257 \\
Response bias \(\beta\) & 1.065 & \multicolumn{2}{c}{---} \\
\bottomrule
\end{tabular}

\footnotesize
\noindent\textit{Note.} \(\beta>1\) indicates a conservative response bias, reflecting a tendency toward ``Human'' attribution.
\end{table}

Overall sensitivity was low and statistically indistinguishable from chance, indicating limited ability to discriminate AI from human teammates.

\paragraph{Detection by task domain}

\begin{table}[ht]
\centering
\caption{\textbf{Identity discrimination by task domain.}}
\label{tab:si_sdt_by_task}
\begin{tabular}{lcccc}
\toprule
\textbf{Task} & \textbf{\(d'\)} & \textbf{95\% CI} & \textbf{Hit rate} & \textbf{False alarm rate} \\
\midrule
Creative story writing (CSW) & 0.220 & [-0.014,\ 0.467] & 0.354 & 0.276 \\
Ethical dilemma (ED) & -0.022 & [-0.252,\ 0.194] & 0.286 & 0.294 \\
Survival ranking (SR) & 0.154 & [-0.089,\ 0.385] & 0.311 & 0.259 \\
\bottomrule
\end{tabular}

\footnotesize
\noindent\textit{Note.} \(d'\) confidence intervals are bootstrap-based. Hit and false-alarm rates are computed using AI and Human judgments only, excluding ``Not sure'' responses.
\end{table}

Across all three task domains, confidence intervals for \(d'\) overlapped zero, indicating no reliable deviation from chance-level identity discrimination as a function of task type.

\paragraph{Detection within mixed-identity triads (clean comparison)}

To isolate the cleanest structural comparison, SDT metrics were estimated within mixed-identity triads (2 humans + 1 AI), where AI and human targets co-occurred under identical group configurations. SDT metrics are undefined in human-only or AI-only triads and are therefore not reported for those conditions.

\begin{table}[ht]
\centering
\caption{\textbf{Identity discrimination within mixed-identity triads (2H+1AI), by AI stance.}}
\label{tab:si_sdt_by_condition_mixed}
\begin{tabular}{lcccc}
\toprule
\textbf{Condition (mixed-identity)} & \textbf{\(d'\)} & \textbf{95\% CI} & \textbf{Hit rate} & \textbf{False alarm rate} \\
\midrule
Contrarian AI (H2\_C) & 0.017 & [-0.255,\ 0.314] & 0.280 & 0.274 \\
Supportive AI (H2\_S) & -0.068 & [-0.368,\ 0.209] & 0.270 & 0.293 \\
\bottomrule
\end{tabular}

\footnotesize
\noindent\textit{Note.} SDT is undefined in AI-only and human-only triads because \(N_{\mathrm{AI}}=0\) or \(N_{\mathrm{H}}=0\).
\end{table}

Within mixed-identity triads, sensitivity remained near zero for both AI stances, with confidence intervals overlapping chance. These results indicate that conversational stance did not improve identity inference when structural conditions were matched.

\paragraph{Participant-level sensitivity}

To examine whether aggregate chance performance masked reliable identity discrimination by a subset of participants, participant-level \(d'\) values were computed based on each participant’s own judgments. Sensitivity estimates clustered tightly around zero (mean \(d'=-0.012\), \(SD=0.697\), 95\% CI \([-0.085,\,0.062]\); \(n=786\)), suggesting uniformly poor identity discrimination at the individual level rather than averaging over heterogeneous strategies.

\begin{table}[ht]
\centering
\caption{\textbf{Participant-level SDT summary.}}
\label{tab:si_sdt_individual}
\begin{tabular}{lcccc}
\toprule
\textbf{Metric} & \textbf{Mean} & \textbf{SD} & \textbf{95\% CI (mean)} & \textbf{\(n\)} \\
\midrule
Participant-level \(d'\) & -0.012 & 0.697 & [-0.085,\ 0.062] & 786 \\
\bottomrule
\end{tabular}
\end{table}

\paragraph{Persona validity (manipulation check among AI targets)}

We next assessed whether participants perceived the intended stance manipulation among AI targets.

\begin{table}[ht]
\centering
\caption{\textbf{Persona validity check: perceived supportiveness and conflictuality by AI stance (AI targets only).}}
\label{tab:si_persona_validity}
\begin{tabular}{lccc}
\toprule
\textbf{Measure} & \textbf{Contrarian AI} \(M\) (SD) & \textbf{Supportive AI} \(M\) (SD) & \textbf{\(n\)} \\
\midrule
Supportiveness & 4.47 (1.91) & 5.77 (1.24) & 341 / 344 \\
Conflictuality & 4.25 (1.85) & 2.51 (1.73) & 341 / 344 \\
\bottomrule
\end{tabular}

\footnotesize
\noindent\textit{Note.} Ratings were provided on a 1--7 scale. Sample sizes are reported as Contrarian / Supportive.
\end{table}

The results confirm that supportive agents were perceived as substantially more supportive and less conflictual than contrarian agents, supporting the validity of the stance manipulation.

\paragraph{Perceptions by ground-truth identity (AI vs human targets)}

Finally, we examined differences in key social perceptions as a function of ground-truth identity.

\begin{table}[ht]
\centering
\caption{\textbf{Perceptions by ground-truth identity (AI vs human targets).}}
\label{tab:si_perceptions_by_truth}
\begin{tabular}{lcccccc}
\toprule
\textbf{Variable} & \textbf{AI Mean} & \textbf{AI 95\% CI} & \textbf{Human Mean} & \textbf{Human 95\% CI} & \textbf{\(t\)} & \textbf{\(p\)} \\
\midrule
Humanness & 5.096 & [4.963,\ 5.230] & 5.398 & [5.295,\ 5.501] & -3.573 & \(<.001\) \\
Trust & 5.223 & [5.106,\ 5.341] & 5.470 & [5.380,\ 5.560] & -3.322 & \(<.001\) \\
Supportiveness & 5.121 & [4.991,\ 5.252] & 5.503 & [5.412,\ 5.593] & -4.862 & \(<.001\) \\
Conflictuality & 3.378 & [3.229,\ 3.528] & 3.034 & [2.913,\ 3.155] & 3.550 & \(<.001\) \\
\bottomrule
\end{tabular}

\footnotesize
\noindent\textit{Note.} Confidence intervals are t-based. \(p\)-values correspond to two-sided independent-samples tests.
\end{table}

Across measures, AI targets were rated slightly lower on humanness and trust than human targets, although effect sizes were modest relative to the extensive identity confusion shown in Table~\ref{tab:si_confusion_matrix}. Together with the persona validity results (Table~\ref{tab:si_persona_validity}), this pattern suggests that participants were sensitive to interactional stance cues and adjusted social evaluations accordingly, even as categorical identity inference remained unreliable.

\begin{table}[ht]
\centering
\caption{\textbf{AI-only perceptions by stance (AI targets only).}}
\label{tab:si_ai_perceptions_by_stance}
\begin{tabular}{lcccc}
\toprule
\textbf{Variable} & \textbf{Contrarian AI} \(M\) (SD) & \textbf{\(n\)} & \textbf{Supportive AI} \(M\) (SD) & \textbf{\(n\)} \\
\midrule
Humanness & 5.035 (1.824) & 341 & 5.157 (1.743) & 344 \\
Trust & 4.871 (1.708) & 341 & 5.573 (1.332) & 344 \\
Supportiveness & 4.466 (1.913) & 341 & 5.770 (1.244) & 344 \\
Conflictuality & 4.252 (1.851) & 341 & 2.512 (1.735) & 344 \\
\bottomrule
\end{tabular}
\end{table}

\subsubsection*{Computational linguistics}
\label{sec:si_impression_texts}

After each identity judgment, participants provided a brief free-response impression text. These texts were analysed to characterise the semantic heuristics participants relied on when forming identity judgments. After excluding empty responses, impression texts were embedded using a pretrained sentence-transformer model (all-MiniLM-L6-v2) and analysed using BERTopic, an unsupervised topic-modelling framework that combines semantic embeddings with density-based clustering. Dimensionality reduction was performed using UMAP and topic clusters were identified using HDBSCAN. Topic representations were derived using class-based term weighting (c-TF--IDF), yielding keyword summaries for each topic.

Topic prevalence was estimated as the proportion of texts assigned to each topic, with uncertainty quantified using 95\% confidence intervals (Wilson score intervals). Documents not assigned to any cluster by HDBSCAN were treated as outliers (topic \(=-1\)) and excluded from inferential analyses. To test whether semantic themes mapped onto identity judgments, we examined associations between topic membership and judgment category using contingency analyses. Effect sizes were quantified using Cramér’s \(V\), with uncertainty estimated via bootstrap confidence intervals. Complementary analyses estimated mutual information (MI) between topic assignments and judgment categories to assess the overall predictive value of semantic themes. To further characterise linguistic contrasts, we conducted supervised class-based TF--IDF analyses contrasting texts associated with different judgment categories, with bootstrap confidence intervals for term weights.

\paragraph{Corpus characteristics and judgment-category composition}

\begin{table}[ht]
\centering
\caption{\textbf{Impression-text corpus characteristics and judgment-category composition.}}
\label{tab:si_text_corpus}
\begin{tabular}{lccc}
\toprule
\textbf{Quantity} & \textbf{Estimate} & \textbf{95\% CI (lower)} & \textbf{95\% CI (upper)} \\
\midrule
Total judgments (rows) & 1{,}572 & \multicolumn{2}{c}{---} \\
Empty/missing texts & 4 (0.3\%) & 0.1\% & 0.7\% \\
Analytic corpus size & 1{,}568 & \multicolumn{2}{c}{---} \\
Mean length (characters) & 66.1 & 63.4 & 68.7 \\
Median length (characters) & 52.0 & \multicolumn{2}{c}{---} \\
Mean word count & 12.2 & 11.7 & 12.7 \\
Median word count & 10.0 & \multicolumn{2}{c}{---} \\
\bottomrule
\end{tabular}

\vspace{6pt}

\begin{tabular}{lrrc}
\toprule
\textbf{Judgment category (after filtering)} & \textbf{\(n\)} & \textbf{\%} & \textbf{95\% CI} \\
\midrule
Human\_Human & 495 & 31.6 & [29.3,\ 33.9] \\
AI\_Human & 358 & 22.8 & [20.8,\ 25.0] \\
Not\_sure & 253 & 16.1 & [14.4,\ 18.0] \\
Human\_AI & 245 & 15.6 & [13.9,\ 17.5] \\
AI\_AI & 217 & 13.8 & [12.2,\ 15.6] \\
\bottomrule
\end{tabular}

\footnotesize
\noindent\textit{Note.} Percentages and CIs are based on the filtered corpus (\(n=1{,}568\)).
\end{table}

\paragraph{BERTopic model configuration and topic prevalence}

UMAP was configured with \(n_{\text{neighbors}}=20\), \(n_{\text{components}}=5\), cosine distance, and a fixed random seed (42). HDBSCAN used minimum cluster size \(=10\) with Euclidean distance. The BERTopic model extracted eight substantive topics and one outlier label (topic \(=-1\)).

\begin{table}[ht]
\centering
\caption{\textbf{Topic prevalence and keyword summaries (BERTopic; \(n=1{,}568\) texts).}}
\label{tab:si_topic_prevalence}
\begin{tabular}{lrrcl}
\toprule
\textbf{Topic} & \textbf{\(n\)} & \textbf{\%} & \textbf{95\% CI} & \textbf{Top keywords (c-TF--IDF)} \\
\midrule
$-1$ (outliers) & 25  & 1.6 & [1.1,\ 2.3] & stuart, bob, really, chat, bots \\
0 & 934 & 59.6 & [57.1,\ 62.0] & way, human, like, responses, answers \\
1 & 177 & 11.3 & [9.8,\ 13.0] & ai, like, human, think, didn \\
2 & 139 & 8.9 & [7.6,\ 10.4] & bob, human, like, responses, ideas \\
3 & 90  & 5.7 & [4.7,\ 7.0] & kevin, ai, human, did, like \\
4 & 88  & 5.6 & [4.6,\ 6.9] & stuart, human, like, ideas, gave \\
5 & 42  & 2.7 & [2.0,\ 3.6] & bob, ai, like, human, opinion \\
6 & 38  & 2.4 & [1.8,\ 3.3] & sure, know, difficult, tell, guess \\
7 & 23  & 1.5 & [1.0,\ 2.2] & stuart, ai, tell, sense, think \\
8 & 12  & 0.8 & [0.4,\ 1.3] & names, real, assigned, introduced, people \\
\bottomrule
\end{tabular}

\footnotesize
\noindent\textit{Note.} Topic prevalence CIs use Wilson score intervals. Topic \(=-1\) denotes outliers (unclustered documents).
\end{table}

\paragraph{Topic similarity and higher-order clustering}

\begin{table}[ht]
\centering
\caption{\textbf{Topic similarity summary (pairwise cosine similarity of topic embeddings).}}
\label{tab:si_topic_similarity}
\begin{tabular}{lccc}
\toprule
\textbf{Statistic} & \textbf{Estimate} & \textbf{95\% CI (lower)} & \textbf{95\% CI (upper)} \\
\midrule
Mean cosine similarity & 0.470 & 0.409 & 0.531 \\
Median cosine similarity & 0.444 & \multicolumn{2}{c}{---} \\
Minimum cosine similarity & 0.226 & \multicolumn{2}{c}{---} \\
Maximum cosine similarity & 0.832 & \multicolumn{2}{c}{---} \\
\bottomrule
\end{tabular}

\footnotesize
\noindent\textit{Note.} High-similarity topic pairs (\(\ge 0.80\)) were Topic 2--5 (0.832) and Topic 4--7 (0.805).
\end{table}

\begin{table}[ht]
\centering
\caption{\textbf{Higher-order clustering of topics (hierarchical clustering; distance threshold = 0.35).}}
\label{tab:si_topic_clusters}
\begin{tabular}{ll}
\toprule
\textbf{Cluster} & \textbf{Topics} \\
\midrule
Cluster 1 & 4, 7 \\
Cluster 2 & 2, 5 \\
Cluster 3 & 0, 1 \\
Cluster 4 & 3 \\
Cluster 5 & 8 \\
Cluster 6 & 6 \\
\bottomrule
\end{tabular}
\end{table}

\paragraph{Topic--judgment associations}

Outliers were excluded from inferential analyses, yielding \(n=1{,}543\) texts with assigned topics. Overall mutual information between topic membership and judgment category was low (MI \(=0.0887\), 95\% CI [0.0794, 0.1212]), indicating limited predictive value of semantic themes for identity judgments.

\begin{table}[ht]
\centering
\caption{\textbf{Topic-by-judgment contingency table (counts; outliers excluded; \(n=1{,}543\)).}}
\label{tab:si_topic_by_judgment_counts}
\begin{tabular}{lrrrrrr}
\toprule
\textbf{Topic} & \textbf{AI\_AI} & \textbf{AI\_Human} & \textbf{Human\_AI} & \textbf{Human\_Human} & \textbf{Not\_sure} & \textbf{Total} \\
\midrule
0 & 126 & 240 & 136 & 341 & 91 & 934 \\
1 & 32 & 23 & 38 & 43 & 41 & 177 \\
2 & 24 & 61 & 11 & 29 & 14 & 139 \\
3 & 6 & 3 & 27 & 25 & 29 & 90 \\
4 & 11 & 11 & 18 & 34 & 14 & 88 \\
5 & 8 & 12 & 5 & 5 & 12 & 42 \\
6 & 2 & 1 & 1 & 1 & 33 & 38 \\
7 & 2 & 3 & 3 & 5 & 10 & 23 \\
8 & 1 & 1 & 2 & 5 & 3 & 12 \\
\midrule
All & 212 & 355 & 241 & 488 & 247 & 1{,}543 \\
\bottomrule
\end{tabular}
\end{table}

\begin{table}[ht]
\centering
\caption{\textbf{Topic--judgment association tests (topic-wise \(\chi^2\) and Cramér’s \(V\)).}}
\label{tab:si_topic_chisq_v}
\begin{tabular}{lccc}
\toprule
\textbf{Topic} & \(\boldsymbol{\chi^2(4)}\) & \textbf{\(p\)} & \textbf{Cramér’s \(V\) [95\% CI]} \\
\midrule
0 & 85.098 & \(<.001\) & 0.235 [0.190,\ 0.289] \\
1 & 25.629 & \(<.001\) & 0.129 [0.089,\ 0.188] \\
2 & 45.089 & \(<.001\) & 0.171 [0.125,\ 0.225] \\
3 & 48.359 & \(<.001\) & 0.177 [0.142,\ 0.234] \\
4 & 7.430  & .115  & 0.069 [0.041,\ 0.124] \\
5 & 11.415 & .022  & 0.086 [0.055,\ 0.142] \\
6 & 145.746 & \(<.001\) & 0.307 [0.243,\ 0.368] \\
7 & 13.267 & .010  & 0.093 [0.043,\ 0.166] \\
8 & 2.394  & .664  & 0.039 [0.027,\ 0.099] \\
\bottomrule
\end{tabular}

\footnotesize
\noindent\textit{Note.} CIs for \(V\) are bootstrap intervals. Topic 6 corresponds to explicit uncertainty statements and showed the strongest association with Not\_sure judgments.
\end{table}

\paragraph{Mutual information (overall and topic-specific)}

\begin{table}[ht]
\centering
\caption{\textbf{Mutual information between topic membership and judgment category.}}
\label{tab:si_topic_mi}
\begin{tabular}{lcc}
\toprule
\textbf{Quantity} & \textbf{MI} & \textbf{95\% CI} \\
\midrule
Overall MI(Topic, Judgment) & 0.0887 & [0.0794,\ 0.1212] \\
\midrule
Topic 0 & 0.0273 & [0.0172,\ 0.0415] \\
Topic 1 & 0.0084 & [0.0040,\ 0.0170] \\
Topic 2 & 0.0136 & [0.0072,\ 0.0249] \\
Topic 3 & 0.0168 & [0.0103,\ 0.0261] \\
Topic 4 & 0.0026 & [0.0010,\ 0.0084] \\
Topic 5 & 0.0039 & [0.0015,\ 0.0107] \\
Topic 6 & 0.0320 & [0.0205,\ 0.0485] \\
Topic 7 & 0.0033 & [0.0010,\ 0.0105] \\
Topic 8 & 0.0009 & [0.0003,\ 0.0051] \\
\bottomrule
\end{tabular}

\footnotesize
\noindent\textit{Note.} CIs are bootstrap intervals (1{,}000 iterations).
\end{table}

\paragraph{Topic-level association with misclassification (odds ratios)}

We quantified whether particular topics were disproportionately associated with misclassification (incorrect vs correct judgments). Odds ratios are reported with 95\% confidence intervals.

\begin{table}[ht]
\centering
\caption{\textbf{Odds ratios for misclassification (incorrect vs correct) by topic.}}
\label{tab:si_topic_or}
\begin{tabular}{lccc}
\toprule
\textbf{Topic} & \textbf{OR} & \textbf{95\% CI} & \textbf{Correct / Incorrect} \\
\midrule
0 & 0.853 & [0.678,\ 1.072] & 467 / 376 \\
1 & 0.950 & [0.665,\ 1.358] & 75 / 61 \\
2 & 1.677 & [1.155,\ 2.436] & 53 / 72 \\
3 & 1.144 & [0.684,\ 1.913] & 31 / 30 \\
4 & 0.744 & [0.461,\ 1.203] & 45 / 29 \\
5 & 1.552 & [0.747,\ 3.221] & 13 / 17 \\
6 & 0.782 & [0.130,\ 4.697] & 3 / 2 \\
7 & 1.007 & [0.336,\ 3.012] & 7 / 6 \\
8 & 0.585 & [0.146,\ 2.350] & 6 / 3 \\
\bottomrule
\end{tabular}

\footnotesize
\noindent\textit{Note.} ``Correct'' and ``Incorrect'' refer to correctness with respect to ground-truth identity; Not\_sure judgments are excluded from this contrast.
\end{table}

\paragraph{Multinomial prediction of judgment category from topic}

As a complementary analysis, we fitted a multinomial logistic regression predicting judgment category from topic membership and estimated bootstrap confidence intervals for predicted class probabilities. Table~\ref{tab:si_topic_multinom_probs} reports the topic-conditional predicted distributions.

\begin{table}[ht]
\centering
\footnotesize
\caption{\textbf{Multinomial predicted probabilities of judgment category given topic (bootstrap 95\% CIs).}}
\label{tab:si_topic_multinom_probs}
\begin{tabular}{lccccc}
\toprule
\textbf{Topic} &
\textbf{P(AI\_AI)} &
\textbf{P(AI\_Human)} &
\textbf{P(Human\_AI)} &
\textbf{P(Human\_Human)} &
\textbf{P(Not\_sure)} \\
\midrule
0 &
0.142 [0.121,\ 0.163] &
0.255 [0.229,\ 0.281] &
0.153 [0.132,\ 0.176] &
0.346 [0.319,\ 0.374] &
0.105 [0.087,\ 0.122] \\
1 &
0.141 [0.123,\ 0.158] &
0.235 [0.215,\ 0.256] &
0.160 [0.141,\ 0.180] &
0.324 [0.302,\ 0.348] &
0.140 [0.122,\ 0.159] \\
2 &
0.139 [0.119,\ 0.159] &
0.214 [0.190,\ 0.238] &
0.165 [0.143,\ 0.186] &
0.299 [0.271,\ 0.328] &
0.183 [0.163,\ 0.205] \\
3 &
0.134 [0.108,\ 0.160] &
0.192 [0.161,\ 0.222] &
0.167 [0.140,\ 0.193] &
0.271 [0.235,\ 0.307] &
0.236 [0.210,\ 0.265] \\
4 &
0.127 [0.095,\ 0.159] &
0.168 [0.132,\ 0.204] &
0.166 [0.132,\ 0.199] &
0.241 [0.197,\ 0.283] &
0.298 [0.259,\ 0.338] \\
5 &
0.117 [0.081,\ 0.156] &
0.144 [0.106,\ 0.186] &
0.161 [0.120,\ 0.202] &
0.210 [0.159,\ 0.259] &
0.368 [0.314,\ 0.424] \\
6 &
0.106 [0.067,\ 0.151] &
0.120 [0.081,\ 0.167] &
0.152 [0.106,\ 0.201] &
0.178 [0.125,\ 0.231] &
0.444 [0.372,\ 0.517] \\
7 &
0.093 [0.054,\ 0.142] &
0.098 [0.061,\ 0.148] &
0.141 [0.091,\ 0.197] &
0.147 [0.094,\ 0.203] &
0.521 [0.432,\ 0.608] \\
8 &
0.080 [0.042,\ 0.131] &
0.078 [0.045,\ 0.128] &
0.127 [0.075,\ 0.189] &
0.118 [0.070,\ 0.176] &
0.597 [0.493,\ 0.692] \\
\bottomrule
\end{tabular}

\footnotesize
\noindent\textit{Note.} Intervals are bootstrap CIs (1{,}000 iterations). Probabilities are topic-conditional and sum to 1 within each topic.
\end{table}

\paragraph{Supervised lexical contrasts (class-based TF--IDF)}

We identified lexically distinctive terms for each judgment category using class-based TF--IDF with bootstrap confidence intervals (1{,}000 iterations). Table~\ref{tab:si_ctfidf_terms} reports the top terms (subset shown for brevity).

\begin{table}[ht]
\centering
\caption{\textbf{Most distinctive terms by judgment category (c-TF--IDF; bootstrap 95\% CIs).}}
\label{tab:si_ctfidf_terms}
\begin{tabular}{llc}
\toprule
\textbf{Category} & \textbf{Term} & \textbf{c-TF--IDF [95\% CI]} \\
\midrule
AI\_AI & fast & 0.391 [0.176,\ 0.451] \\
AI\_AI & disagree & 0.177 [0.045,\ 0.260] \\
AI\_AI & quick & 0.164 [0.051,\ 0.235] \\
\midrule
AI\_Human & interacted & 0.214 [0.084,\ 0.291] \\
AI\_Human & slang & 0.180 [0.065,\ 0.236] \\
AI\_Human & emotions & 0.162 [0.059,\ 0.226] \\
\midrule
Human\_AI & fast & 0.216 [0.080,\ 0.272] \\
Human\_AI & repeating & 0.140 [0.023,\ 0.210] \\
Human\_AI & script & 0.104 [0.000,\ 0.191] \\
\midrule
Human\_Human & reasoning & 0.257 [0.121,\ 0.317] \\
Human\_Human & normal & 0.225 [0.087,\ 0.321] \\
Human\_Human & engaging & 0.170 [0.063,\ 0.235] \\
\midrule
Not\_sure & tell & 0.383 [0.196,\ 0.453] \\
Not\_sure & know & 0.340 [0.179,\ 0.396] \\
Not\_sure & difficult & 0.170 [0.044,\ 0.250] \\
\bottomrule
\end{tabular}

\footnotesize
\noindent\textit{Note.} This table reports a compact subset of the top-ranked terms per class (see the exported CSV for full lists).
\end{table}

\subsubsection*{Predictive modeling}
This subsection reports the full statistical results underlying the contrast between models predicting participants’ identity judgments and models predicting ground-truth identity from the same dialogue-derived cues. All predictors were standardised prior to modelling. Judgment models used participant-level cluster-robust standard errors to account for repeated teammate evaluations by the same participant, whereas ground-truth models used group-level clustering to provide conservative inference under nested interaction structure. All confidence intervals are 95\%.

\paragraph{Analytic samples and model overview}
We estimated two judgment models to isolate directional attribution (AI vs Human) from explicit uncertainty reporting (Not sure vs Human), using a common human-judgment reference category to enable direct comparison of cue profiles. To assess cue diagnosticity independent of human belief, we estimated ground-truth models using the same standardised predictors. Because experimental condition quasi-deterministically encodes identity in one-human and all-human compositions, the primary truth model was restricted to mixed-identity H2 groups (2 humans + 1 AI), where AI and human targets co-occur under matched interaction structure; the full-sample truth model is reported as a robustness check with maximal power.

\begin{table}[ht]
\centering
\caption{\textbf{Analytic samples and model specifications for judgment and ground-truth prediction.}}
\label{tab:si_jt_samples}
\begin{tabular}{lcccc}
\toprule
\textbf{Model} & \textbf{Outcome} & \textbf{Sample} & \textbf{\(N\)} & \textbf{Clustering} \\
\midrule
Judgment model 1 & AI vs Human & Excluding Not sure & 1{,}282 & Participant \\
Judgment model 2 & Not sure vs Human & Excluding AI & 1{,}072 & Participant \\
Truth model (primary) & AI vs Human (truth) & H2 only (2H+1AI) & 689 & Group \\
Truth model (robustness) & AI vs Human (truth) & Full sample & 1{,}517 & Group \\
\bottomrule
\end{tabular}
\end{table}

\paragraph{Predicting identity judgments}
These models test whether participants’ categorical attributions are systematically coupled to interactional cues beyond task type and group condition. Coefficients are standardised log-odds effects: positive values indicate higher odds of AI attribution (or Not sure, respectively), holding all other cues and covariates constant. Overall, cue-based prediction of judgments was statistically detectable but weak, consistent with limited reliance on these cues in explicit identity inference once structural context was accounted for.

\begin{table}[ht]
\centering
\caption{\textbf{Logistic regression predicting AI versus Human judgments (excluding Not sure).}}
\label{tab:si_judgment_ai_human}
\begin{tabular}{lccc}
\toprule
\textbf{Predictor} & \(\boldsymbol{\beta}\) & \textbf{SE} & \textbf{95\% CI} \\
\midrule
Authenticity & 0.101 & 0.072 & [-0.041,\ 0.242] \\
Function-word rate & -0.044 & 0.090 & [-0.221,\ 0.134] \\
Lexical diversity & -0.005 & 0.069 & [-0.139,\ 0.130] \\
Affect density & -0.035 & 0.078 & [-0.188,\ 0.117] \\
Tone score & 0.107 & 0.082 & [-0.054,\ 0.268] \\
Mean latency & -0.210 & 0.094 & [-0.394,\ -0.027] \\
Latency variability & 0.046 & 0.085 & [-0.121,\ 0.212] \\
Negation rate & 0.137 & 0.076 & [-0.012,\ 0.286] \\
Analytic style & 0.068 & 0.074 & [-0.076,\ 0.213] \\
Conversationality & -0.242 & 0.100 & [-0.438,\ -0.046] \\
\midrule
Pseudo-\(R^2\) & \multicolumn{3}{c}{0.018} \\
LR test \(P\) & \multicolumn{3}{c}{.023} \\
\bottomrule
\end{tabular}
\end{table}

The uncertainty model captures a distinct response behaviour: rather than directional classification, participants explicitly indicated indeterminacy. As such, coefficients should be interpreted as cues associated with non-commitment to either category (Not sure) relative to confident Human judgments. The model provides little evidence that the aggregated cue set strongly structures uncertainty reporting.

\begin{table}[ht]
\centering
\caption{\textbf{Logistic regression predicting Not sure versus Human judgments (excluding AI).}}
\label{tab:si_judgment_notsure}
\begin{tabular}{lccc}
\toprule
\textbf{Predictor} & \(\boldsymbol{\beta}\) & \textbf{SE} & \textbf{95\% CI} \\
\midrule
Authenticity & 0.199 & 0.089 & [0.025,\ 0.374] \\
Function-word rate & -0.037 & 0.122 & [-0.276,\ 0.201] \\
Lexical diversity & -0.052 & 0.090 & [-0.227,\ 0.124] \\
Affect density & 0.081 & 0.091 & [-0.097,\ 0.259] \\
Tone score & 0.093 & 0.100 & [-0.103,\ 0.288] \\
Mean latency & -0.080 & 0.104 & [-0.284,\ 0.124] \\
Latency variability & 0.057 & 0.105 & [-0.149,\ 0.264] \\
Negation rate & 0.115 & 0.091 & [-0.063,\ 0.294] \\
Analytic style & 0.143 & 0.093 & [-0.038,\ 0.325] \\
Conversationality & -0.223 & 0.114 & [-0.445,\ 0.000] \\
\midrule
Pseudo-\(R^2\) & \multicolumn{3}{c}{0.015} \\
LR test \(P\) & \multicolumn{3}{c}{.361} \\
\bottomrule
\end{tabular}
\end{table}

\paragraph{Predicting ground-truth identity}
Ground-truth models quantify whether the same interactional cues are diagnostically informative about actual identity, independent of participants’ beliefs. The primary specification restricts inference to H2 mixed-identity groups to avoid quasi-separation by condition and to ensure a within-structure test where AI and human targets appear in the same interaction context. Coefficients are standardised log-odds effects on the probability that a target is AI.

\begin{table}[ht]
\centering
\caption{\textbf{Logistic regression predicting ground-truth AI identity in mixed-identity groups (H2 only; primary model).}}
\label{tab:si_truth_h2}
\begin{tabular}{lccc}
\toprule
\textbf{Predictor} & \(\boldsymbol{\beta}\) & \textbf{SE} & \textbf{95\% CI} \\
\midrule
Authenticity & 0.124 & 0.249 & [-0.365,\ 0.612] \\
Function-word rate & -1.401 & 0.294 & [-1.978,\ -0.824] \\
Lexical diversity & 0.380 & 0.160 & [0.067,\ 0.693] \\
Affect density & 0.007 & 0.221 & [-0.427,\ 0.441] \\
Tone score & 1.134 & 0.400 & [0.350,\ 1.917] \\
Mean latency & -0.054 & 0.330 & [-0.701,\ 0.594] \\
Latency variability & 0.325 & 0.285 & [-0.234,\ 0.884] \\
Negation rate & 0.978 & 0.835 & [-0.659,\ 2.615] \\
Analytic style & 0.240 & 0.254 & [-0.258,\ 0.738] \\
Conversationality & 2.380 & 0.607 & [1.190,\ 3.570] \\
\midrule
Pseudo-\(R^2\) & \multicolumn{3}{c}{0.692} \\
LR test \(P\) & \multicolumn{3}{c}{\(<.001\)} \\
\bottomrule
\end{tabular}
\end{table}

The full-sample truth model is reported as a robustness check to show that coefficient directions persist under maximal power. Because this specification aggregates across structurally different conditions, it is not used as the primary diagnosticity estimate, but it corroborates that multiple cues differentiate AI from human dialogue at scale.

\begin{table}[ht]
\centering
\caption{\textbf{Ground-truth prediction on the full sample (robustness check).}}
\label{tab:si_truth_full}
\begin{tabular}{lccc}
\toprule
\textbf{Predictor} & \(\boldsymbol{\beta}\) & \textbf{SE} & \textbf{95\% CI} \\
\midrule
Authenticity & 0.461 & 0.158 & [0.150,\ 0.772] \\
Function-word rate & -1.474 & 0.283 & [-2.030,\ -0.919] \\
Lexical diversity & 0.463 & 0.171 & [0.128,\ 0.798] \\
Tone score & 1.320 & 0.275 & [0.781,\ 1.859] \\
Mean latency & 0.605 & 0.206 & [0.201,\ 1.010] \\
Negation rate & 1.159 & 0.437 & [0.302,\ 2.017] \\
Conversationality & 2.199 & 0.526 & [1.167,\ 3.230] \\
\midrule
Pseudo-\(R^2\) & \multicolumn{3}{c}{0.716} \\
\bottomrule
\end{tabular}
\end{table}

\paragraph{Model comparison}
To directly quantify judgment--truth dissociation, we summarised fit statistics across models. Judgment models exhibited minimal explanatory power (pseudo-\(R^2\approx .02\)), whereas truth models achieved large pseudo-\(R^2\) values (\(\approx .69\)--.72), indicating that the cue set contains substantial identity information that participants’ judgments largely fail to incorporate. Because pseudo-\(R^2\) values are not comparable across different outcome constructions in a strict likelihood sense, this comparison is interpreted descriptively as an index of signal strength captured by each modelling target under consistent predictors and covariate adjustment.

\begin{table}[ht]
\centering
\caption{\textbf{Comparison of judgment and truth models.}}
\label{tab:si_jt_comparison}
\begin{tabular}{lcccc}
\toprule
\textbf{Model} & \textbf{Pseudo-\(R^2\)} & \textbf{LogLik} & \textbf{AIC} & \textbf{\(N\)} \\
\midrule
AI vs Human judgment & 0.018 & -813.15 & 1660.29 & 1{,}282 \\
Not sure vs Human judgment & 0.015 & -555.09 & 1144.18 & 1{,}072 \\
Ground truth (H2 only) & 0.692 & -147.23 & 320.46 & 689 \\
Ground truth (full sample) & 0.716 & -295.34 & 616.69 & 1{,}517 \\
\bottomrule
\end{tabular}
\end{table}

\subsubsection*{Robustness}
\label{si:robustness}

\paragraph{Robustness checks for judgment models}

We first examined whether inference for the judgment models was sensitive to the level at which cluster-robust standard errors were computed. The primary specification clustered standard errors at the participant level to account for repeated judgments by the same individual. As a robustness check, we re-estimated the same judgment models using group-level clustering. Coefficient signs, magnitudes, and substantive conclusions were unchanged under group-level clustering, indicating that the observed weak coupling between dialogue cues and identity judgments did not depend on the choice of clustering level.

We additionally assessed multicollinearity among dialogue-derived predictors using variance inflation factors (VIFs). All VIF values were well below conventional thresholds, indicating no problematic collinearity. Finally, to test sensitivity to low-information targets, we re-estimated the AI-versus-Human judgment model after excluding the bottom decile of targets by total word count (threshold = 67 words), which removed 9.8\% of judgments. Results remained qualitatively unchanged.

\begin{table}[ht]
\centering
\caption{\textbf{Collinearity diagnostics for dialogue-derived predictors.} Variance inflation factors (VIFs) computed on the full analytic dataset (\(N=1{,}517\)). Values above 10 are commonly interpreted as indicating problematic collinearity.}
\label{tab:si_vif}
\begin{tabular}{lc}
\toprule
\textbf{Predictor} & \textbf{VIF} \\
\midrule
Authenticity & 1.120 \\
Function-word rate & 1.875 \\
Lexical diversity & 1.238 \\
Affect density & 1.427 \\
Tone score & 1.463 \\
Mean latency & 1.974 \\
Latency variability & 1.989 \\
Negation rate & 1.343 \\
Analytic style & 1.361 \\
Conversationality & 1.820 \\
\bottomrule
\end{tabular}
\end{table}

\paragraph{Target-level robustness: within-triad identification of AI}

To determine whether cue diagnosticity reflected within-group identity information rather than between-group confounding, we conducted robustness analyses at the \emph{target level}, with one row per evaluated teammate. The analysis was restricted to \emph{complete-evaluation H2 triads} (2 humans + 1 AI) in which all three teammates received at least one human evaluation. This restriction matches the experimental design—AI targets are rated by humans but do not themselves provide ratings—and ensures valid within-triad identification. The resulting analytic universe comprised \(N=149\) triads and \(N=447\) evaluated targets (149 AI; 298 human).

We initially attempted a conditional logistic regression stratified by group to identify diagnosticity purely from within-group variation. Because the conditional likelihood implementation did not yield an estimable covariance matrix for uncertainty quantification, we instead implemented a mathematically equivalent group fixed-effects logistic regression with group indicators. This specification controls for all group-level confounds and identifies AI targets solely via within-group differences.

\begin{table}[ht]
\centering
\caption{\textbf{Within-triad diagnosticity using group fixed-effects logistic regression (target-level; complete-evaluation H2 triads).} Coefficients are standardised log-odds. Odds ratios (ORs) are reported for interpretability. Task type was included, and group fixed effects absorb all between-group differences.}
\label{tab:si_group_fe_truth}
\begin{tabular}{lcccccc}
\toprule
\textbf{Predictor} & \(\boldsymbol{\beta}\) & \textbf{SE} & \textbf{\(z\)} & \textbf{\(P\)} & \textbf{95\% CI} & \textbf{OR [95\% CI]} \\
\midrule
Authenticity & -0.108 & 0.586 & -0.184 & .854 & [-1.257,\ 1.041] & 0.898 [0.284,\ 2.832] \\
Function-word rate & -2.155 & 0.949 & -2.270 & .023 & [-4.015,\ -0.295] & 0.116 [0.018,\ 0.745] \\
Lexical diversity & 0.676 & 0.680 & 0.994 & .320 & [-0.657,\ 2.010] & 1.966 [0.518,\ 7.460] \\
Affect density & -0.451 & 0.837 & -0.538 & .590 & [-2.091,\ 1.190] & 0.637 [0.124,\ 3.286] \\
Tone score & 3.194 & 0.917 & 3.484 & \(<.001\) & [1.397,\ 4.992] & 24.397 [4.044,\ 147.179] \\
\bottomrule
\end{tabular}
\end{table}

\paragraph{Out-of-sample discrimination and calibration}

We next evaluated robustness under strict out-of-sample validation using group-wise cross-validation, holding out entire groups in each fold. Performance was summarised using AUC (discrimination), Brier score (probability accuracy), and calibration diagnostics (slope, intercept, and expected calibration error; ECE). The same group-wise validation protocol was applied to (i) the target-level ground-truth model in the complete-evaluation H2 universe and (ii) the judgment-level models (AI vs Human; Not sure vs Human). This allows a direct comparison of how well dialogue cues predict ground-truth identity versus how well they predict human judgments under identical validation structure.

\begin{table}[ht]
\centering
\scriptsize
\caption{\textbf{Group-wise cross-validation performance for truth versus judgment prediction.} Values are mean \(\pm\) SD across folds. Ideal calibration corresponds to slope = 1 and intercept = 0; lower ECE indicates better calibration.}
\label{tab:si_cv_metrics}
\begin{tabular}{lcccccc}
\toprule
\textbf{Model} & \textbf{Unit} & \textbf{\(N\)} & \textbf{Groups} & \textbf{AUC} & \textbf{Brier} & \textbf{Calibration (slope, intercept, ECE)} \\
\midrule
Truth model (H2 target-level) & Target & 447 & 149 & 0.983 \(\pm\) 0.015 & 0.046 \(\pm\) 0.017 & 0.746,\ -0.075,\ 0.028 \\
Judgment model (AI vs Human) & Judgment & 1{,}282 & 449 & 0.527 \(\pm\) 0.025 & 0.228 \(\pm\) 0.008 & 0.338,\ -0.412,\ 0.047 \\
Judgment model (Not sure vs Human) & Judgment & 1{,}072 & 413 & 0.500 \(\pm\) 0.028 & 0.175 \(\pm\) 0.009 & 0.004,\ -1.265,\ 0.056 \\
\bottomrule
\end{tabular}
\end{table}

As an additional sensitivity check, we examined calibration under an in-sample fit of the target-level truth model (complete-evaluation H2 universe). In-sample estimation yielded slightly optimistic performance and improved calibration relative to cross-validation (AUC = 0.988; Brier = 0.033; slope = 1.000; intercept = 0.001; ECE = 0.019), indicating that the reliability pattern observed under cross-validation was not driven by fold-specific artefacts.

\paragraph{Timing ablation: excluding latency features}

To test whether diagnosticity was driven primarily by timing cues, we repeated group-wise cross-validation for the target-level truth model after excluding the two latency-related predictors (mean latency and latency variability), retaining the remaining eight predictors. Discrimination remained near ceiling, with virtually no change in AUC, and both Brier score and calibration improved slightly.

\begin{table}[ht]
\centering
\footnotesize
\caption{\textbf{Timing ablation for target-level ground-truth prediction (complete-evaluation H2 triads).} Metrics are group-wise cross-validation mean \(\pm\) SD across folds. \(\Delta\)AUC is computed as Full minus No-timing.}
\label{tab:si_timing_ablation}
\begin{tabular}{lcccc}
\toprule
\textbf{Specification} & \textbf{AUC} & \textbf{Brier} & \textbf{Calibration (slope, intercept, ECE)} & \(\boldsymbol{\Delta}\)\textbf{AUC} \\
\midrule
All features (including timing) & 0.983 \(\pm\) 0.015 & 0.046 \(\pm\) 0.017 & 0.746,\ -0.075,\ 0.028 & -- \\
Excluding latency features & 0.983 \(\pm\) 0.015 & 0.042 \(\pm\) 0.017 & 0.807,\ -0.059,\ 0.019 & -0.001 \\
\bottomrule
\end{tabular}
\end{table}

\paragraph{Permutation test preserving within-triad structure}

To confirm that the high discrimination observed for the target-level model was not an artefact of the cross-validation procedure or class imbalance, we conducted a within-triad permutation test that preserves the mixed-identity structure. In each permutation, exactly one AI label was randomly assigned per triad at the target level, AUC was recomputed under the same evaluation pipeline, and the procedure was repeated 1{,}000 times to generate a null distribution. The observed AUC lay far in the upper tail of this null distribution, yielding a permutation \(p<.001\).

\begin{table}[ht]
\centering
\caption{\textbf{Permutation test for target-level diagnosticity (complete-evaluation H2 triads).} The null distribution preserves triad structure by enforcing exactly one AI label per triad.}
\label{tab:si_permutation}
\begin{tabular}{lc}
\toprule
\textbf{Quantity} & \textbf{Value} \\
\midrule
Observed AUC & 0.9825 \\
Null AUC mean \(\pm\) SD & 0.5845 \(\pm\) 0.0208 \\
Null median AUC & 0.5839 \\
Null 95\% interval & [0.5466,\ 0.6292] \\
Permutations & 1{,}000 \\
Permutation \(p\) & \(<.001\) \\
\bottomrule
\end{tabular}
\end{table}

\paragraph{Triad-level Top-1 identification}

Finally, we evaluated a triad-level decision rule aligned with the experimental constraint that each H2 triad contains exactly one AI teammate. Within each triad, the teammate with the highest predicted AI probability was selected, and this Top-1 selection was scored against the true AI target. Top-1 accuracy was far above the 1/3 chance baseline and remained comparably high when latency features were excluded.

\begin{table}[ht]
\centering
\caption{\textbf{Top-1 triad identification accuracy for selecting the AI teammate within complete-evaluation H2 triads.} Confidence intervals are bootstrap 95\% CIs across triads.}
\label{tab:si_top1}
\begin{tabular}{lcccc}
\toprule
\textbf{Specification} & \textbf{Triads} & \textbf{Top-1 accuracy} & \textbf{95\% CI} & \textbf{Chance} \\
\midrule
All features (including timing) & 149 & 0.953 & [0.906,\ 1.000] & 0.333 \\
Excluding latency features & 149 & 0.966 & [0.924,\ 1.000] & 0.333 \\
\bottomrule
\end{tabular}
\end{table}

\subsubsection*{Representational similarity}
\label{si:rsa}

\paragraph{Overview and target-level aggregation}

We conducted representational similarity analysis (RSA) to examine whether the relational geometry among evaluated teammates is shared across multiple representational spaces: (i) interactional dialogue cues, (ii) participants’ identity judgments, (iii) ground-truth identity, (iv) subjective impressions, and (v) semantic topics derived from impression narratives. The unit of analysis was the \emph{target}, defined as a unique teammate within a collaborative group.

Judgment-level observations were aggregated to the target level by computing, for each target, the modal identity judgment (Human, AI, or Not sure) across all evaluations received. In addition, each target was characterised by its ground-truth identity, mean interactional cue values, mean impression ratings (humanness and trust), and the dominant BERTopic topic associated with its impression text. Targets with incomplete cue data were excluded, yielding \(N=1{,}148\) targets with complete interactional and judgment profiles. For topic-based RSA, targets without valid topic assignments were excluded, resulting in \(N=1{,}147\) targets for topic RDMs. For the most stringent mixed-identity test, analyses were further restricted to H2 groups (2 humans + 1 AI), yielding \(N=539\) targets.

\begin{table}[ht]
\centering
\caption{\textbf{Target-level sample composition for RSA.} Targets are unique teammates within groups. Modal judgments are computed across all evaluations received by a target.}
\label{tab:si_rsa_sample}
\begin{tabular}{lccc}
\toprule
\textbf{Quantity} & \textbf{Full sample} & \textbf{H2-only} & \textbf{Notes} \\
\midrule
Targets (complete cue data) & 1{,}148 & 539 & Interactional and judgment RSA \\
Targets with valid topic assignment & 1{,}147 & -- & Topic RSA excludes missing topics \\
Modal judgments: Human & 612 & -- & Target-level mode across raters \\
Modal judgments: AI & 417 & -- & Target-level mode across raters \\
Modal judgments: Not sure & 119 & -- & Target-level mode across raters \\
Ground truth: Human targets & 632 & -- & Target-level ground truth \\
Ground truth: AI targets & 516 & -- & Target-level ground truth \\
\bottomrule
\end{tabular}
\end{table}

\paragraph{Representational dissimilarity matrices}

For each representational space, we constructed a representational dissimilarity matrix (RDM) over targets. Interactional cue and impression RDMs were computed as cosine distances over standardised feature vectors (10 interactional cues; 2 impression dimensions: humanness and trust). The ground-truth RDM encoded binary dissimilarity (0 for the same class; 1 for different classes). The topic RDM encoded binary dissimilarity based on whether two targets shared the same dominant topic. The judgment RDM encoded a graded dissimilarity that reflected uncertainty, assigning intermediate dissimilarity to pairs involving Not sure (with exact coding fixed a priori in the analysis script).

RSA correlations were computed as Spearman correlations between the upper triangles of RDM pairs. Uncertainty was quantified using bootstrap confidence intervals obtained by resampling dissimilarity pairs (1{,}000 iterations). Statistical significance was assessed using permutation tests that permuted target labels (1{,}000 iterations), recomputing RSA correlations to generate null distributions.

\paragraph{Primary RSA results}

Across the full target set, the geometry of interactional cues aligned strongly with ground-truth identity (\(\rho=0.455\), 95\% CI [0.453, 0.456], permutation \(p<.001\)), indicating that AI and human targets occupied separable regions in cue space. In contrast, cue geometry was not reflected in participants’ modal identity judgments (\(\rho\approx0\), 95\% CI [-0.002, 0.002], permutation \(p=.924\)), and judgment structure showed only negligible alignment with ground-truth identity (\(\rho=0.004\), 95\% CI [0.002, 0.007], permutation \(p=.027\)).

Instead, judgment structure aligned with subjective impression space defined by humanness and trust (\(\rho=0.247\), 95\% CI [0.245, 0.250], permutation \(p<.001\)), whereas impression structure was effectively orthogonal to cue structure (\(\rho=0.007\), 95\% CI [0.004, 0.009], permutation \(p<.001\)). Topic structure exhibited negligible correspondence with judgment space (\(\rho=0.025\), 95\% CI [0.022, 0.027], permutation \(p<.001\)) and no correspondence with cue space (\(\rho\approx0\), 95\% CI [-0.003, 0.002], permutation \(p=.806\)).

\begin{table}[ht]
\centering
\caption{\textbf{Representational similarity analysis (RSA) correlations across representational spaces (full sample).} Spearman correlations are computed between the upper triangles of each pair of RDMs. Confidence intervals are 95\% bootstrap CIs; permutation \(p\)-values are based on 1{,}000 label permutations.}
\label{tab:si_rsa_full}
\begin{tabular}{lccccc}
\toprule
\textbf{RDM comparison} & \(\boldsymbol{\rho}\) & \textbf{95\% CI} & \textbf{Effect} & \textbf{Permutation \(p\)} & \textbf{Pairs} \\
\midrule
Cue \(\leftrightarrow\) Ground truth & 0.455 & [0.453,\ 0.456] & Medium & \(<.001\) & 658{,}378 \\
Cue \(\leftrightarrow\) Judgment & -0.000 & [-0.002,\ 0.002] & Negligible & .924 & 658{,}378 \\
Judgment \(\leftrightarrow\) Ground truth & 0.004 & [0.002,\ 0.007] & Negligible & .027 & 658{,}378 \\
Impression \(\leftrightarrow\) Judgment & 0.247 & [0.245,\ 0.250] & Small & \(<.001\) & 658{,}378 \\
Cue \(\leftrightarrow\) Impression & 0.007 & [0.004,\ 0.009] & Negligible & \(<.001\) & 658{,}378 \\
Topic \(\leftrightarrow\) Judgment & 0.025 & [0.022,\ 0.027] & Negligible & \(<.001\) & -- \\
Topic \(\leftrightarrow\) Cue & -0.000 & [-0.003,\ 0.002] & Negligible & .806 & -- \\
\bottomrule
\end{tabular}
\end{table}

\paragraph{Stringent mixed-identity (H2-only) RSA}

To provide a stringent test in which AI and human targets co-occur under identical structural conditions, we repeated RSA within H2 groups (2 humans + 1 AI). Under this restriction, cue geometry remained strongly aligned with ground-truth identity (\(\rho=0.415\), 95\% CI [0.410, 0.419], permutation \(p<.001\)), while cue--judgment alignment remained negligible (\(\rho=0.008\), 95\% CI [0.003, 0.013], permutation \(p=.014\)). Alignment between judgment structure and ground truth increased slightly but remained small in magnitude (\(\rho=0.031\), 95\% CI [0.026, 0.037], permutation \(p<.001\)). Together, these results demonstrate that the representational dissociation persists even when interaction context is held constant.

\begin{table}[ht]
\centering
\caption{\textbf{RSA within mixed-identity triads (H2-only).} Spearman correlations between RDMs computed over targets in H2 groups only. Confidence intervals are 95\% bootstrap CIs; permutation \(p\)-values are based on 1{,}000 label permutations.}
\label{tab:si_rsa_h2}
\begin{tabular}{lccccc}
\toprule
\textbf{RDM comparison (H2)} & \(\boldsymbol{\rho}\) & \textbf{95\% CI} & \textbf{Effect} & \textbf{Permutation \(p\)} & \textbf{Pairs} \\
\midrule
Cue \(\leftrightarrow\) Ground truth & 0.415 & [0.410,\ 0.419] & Medium & \(<.001\) & 144{,}991 \\
Cue \(\leftrightarrow\) Judgment & 0.008 & [0.003,\ 0.013] & Negligible & .014 & 144{,}991 \\
Judgment \(\leftrightarrow\) Ground truth & 0.031 & [0.026,\ 0.037] & Negligible & \(<.001\) & 144{,}991 \\
\bottomrule
\end{tabular}
\end{table}

\end{document}